\documentclass[journal]{IEEEtran}

\IEEEoverridecommandlockouts

\usepackage[utf8]{inputenc}
\usepackage[T1]{fontenc}
\usepackage{url}
\usepackage{booktabs}
\usepackage{tabularx}
\usepackage{amsfonts}
\usepackage{nicefrac}
\usepackage{microtype}
\usepackage{xcolor}
\usepackage[cmex10]{amsmath}
\usepackage{cuted}

\usepackage{graphicx}
\usepackage[labelsep=period]{caption}
\usepackage{subcaption}
\usepackage{makecell}
\usepackage{amsmath,amsfonts,amssymb,amsthm,epstopdf,url,array}
\usepackage{cases}
\usepackage{bm}
\usepackage{bbm}
\usepackage{cite}
\usepackage[ruled,linesnumbered]{algorithm2e}
\SetKwInput{KwInput}{Input}
\SetKwInput{KwOutput}{Output}
\SetKwComment{Comment}{// }{}
\SetKw{Return}{return}

\usepackage{xparse}
\usepackage{mathtools}
\usepackage{multirow}
\usepackage{xr-hyper}
\usepackage[hidelinks]{hyperref}

\usepackage[flushleft]{threeparttable}

\newtheorem{lemma}{Lemma}
\newtheorem{proposition}{Proposition}

\theoremstyle{definition}
\newtheorem{definition}{Definition}

\theoremstyle{remark}

\theoremstyle{plain}

\DeclareMathOperator*{\argmax}{arg\,max}
\DeclareMathOperator*{\argmin}{arg\,min}
\DeclareMathOperator{\diag}{diag}
\DeclareMathOperator{\Cov}{Cov}

\newcommand{\E}{\mathbb{E}}
\newcommand{\Normal}{\mathcal{N}}
\newcommand{\Ind}{\mathbbm{1}}

\newcommand{\Pone}{\textup{\textbf{P1}}}
\newcommand{\Ponep}{\textup{\textbf{P1}}\ensuremath{'}}
\newcommand{\Ptwo}{\textup{\textbf{P2}}}

\begin{document}

\title{Task-Driven Maintenance of Dynamic Channel Knowledge Maps in LEO-Assisted Edge Networks}

\author{
Linghao~Zhang,
Zixiang~Ren,
Haitao~Zhao,~\IEEEmembership{Senior~Member,~IEEE,}
Bo~Xu,
and~Rui~Zhang,~\IEEEmembership{Fellow,~IEEE}

\thanks{This work was supported by the National Natural Science Foundation of China under Grant U24B20187 and Grant 62371250, the Major Science and Technology Project of Jiangsu Province under Grant BG2024002, and the Postgraduate Research \& Practice Innovation Program of Jiangsu Province under Grant KYCX24\_1197. \textit{(Corresponding author: Haitao Zhao.)}}

\thanks{Linghao Zhang, Haitao Zhao, and Bo Xu are with the Jiangsu Key Laboratory of New-Generation Mobile Communications, Nanjing University of Posts and Telecommunications, Nanjing 210003, China (e-mail: \mbox{zhanglinghao@njupt.edu.cn}; \mbox{zhaoht@njupt.edu.cn}; \mbox{xubo@njupt.edu.cn}).}

\thanks{Zixiang Ren and Rui Zhang are with the Department of Electrical and Computer Engineering, National University of Singapore, Singapore (e-mail: \mbox{zixiang\_ren@nus.edu.sg}; \mbox{elezhang@nus.edu.sg}).}
}

\maketitle

\bstctlcite{IEEEexample:BSTcontrol}

\begin{abstract}
In low Earth orbit (LEO) satellite-assisted edge networks, network-level decisions such as user association and computation offloading query a channel knowledge map (CKM) for large-scale channel knowledge. Yet this knowledge ages within minutes, and refreshing it beyond the routine reports of connected users requires costly extra measurements. Most existing methods maintain a CKM by its error, uncertainty, or age, whereas a map error matters only when it alters a decision. We therefore propose to maintain the CKM by how much each measurement improves the decisions that query it. Specifically, we model the CKM as a Gaussian belief over environment residuals indexed by geographic cell, base station, and satellite sky direction, which ages between reports and updates via Kalman filtering. Because both the decision regret and its expected reduction from a prospective measurement can be evaluated prior to measurement, we formulate CKM maintenance as minimizing the long-run decision regret subject to a measurement budget. To solve this problem, we develop task-driven CKM maintenance (TD-CKM), an efficient algorithm that evaluates the expected regret reduction in closed form across current and upcoming decisions, along with an adaptive Lagrangian pricing mechanism that satisfies the long-run budget on every sample path. City-scale simulations demonstrate that TD-CKM matches the regret of the strongest baseline using 30\% to 40\% fewer extra measurements. Compared with relying on routine reports alone, map-centric policies achieve a larger reduction in map error than TD-CKM, yet reduce decision regret by less than 4\%, compared with 36\% under TD-CKM.
\end{abstract}


\section{Introduction}\label{sec:intro}

Integrating low Earth orbit (LEO) satellites into terrestrial edge networks allows users to access communication and computing services from both base stations (BSs) and satellites~\cite{azari2022evolution}. Operating such LEO-assisted edge networks relies on network-level decisions, including user association, handover, and computation offloading, which evaluate terrestrial and satellite candidates over time scales of seconds to minutes~\cite{zhang2026tvt}. Among these, user association and partial offloading typify discrete and continuous decisions, respectively. Over such time scales, small-scale fading averages out, leaving these decisions dependent on large-scale channel information such as the channel gain and average signal-to-noise ratio (SNR) of each candidate link. In particular, user association requires this information even for candidate BSs and satellites not currently serving the user. Acquiring it via dedicated measurements incurs measurement gaps that interrupt ongoing data transmission, while consuming terminal energy and signaling overhead. Nevertheless, most decision designs assume that the large-scale channel information of all candidates is readily available, without accounting for how the network acquires it~\cite{nguyenkha2024twotier}.

A channel knowledge map (CKM) is a site-specific database that stores location-indexed channel information, such as the channel gain between each location and network node, to reduce or eliminate real-time channel measurement~\cite{zeng2024tutorial}. Constructed from channel measurements reported by users along with their locations, a CKM completes unmeasured locations through spatial interpolation, machine learning, or physical environment modeling~\cite{ren2026ckmsurvey}. Once constructed, the stored knowledge can be reused across users and over time. In an LEO-assisted edge network, a CKM enables network-level decisions to directly query the large-scale channel knowledge of all terrestrial and satellite candidates rather than measuring each link individually, and knowledge indexed by sky direction can be reused by subsequent satellites passing through the same direction.

In a dynamic network, however, stored channel knowledge quickly loses both accuracy and relevance. Its accuracy degrades over minutes to hours as temporary blockages, foliage, moving vehicles, and rain alter propagation conditions. Meanwhile, its relevance shifts as user mobility redistributes the demand for channel knowledge and as satellite motion predictably changes which sky directions are queried by decisions. Although routine reports from connected users refresh part of the map at no extra cost, these reports are sent only periodically to limit uplink and backhaul signaling. Between routine reports and for the links they do not cover, updating the CKM requires extra measurements, which reintroduce the acquisition costs that the map is meant to avoid. While a service region contains thousands of candidate links whose channel knowledge ages within minutes, the network can afford to measure only a small fraction of them in each decision slot. Maintaining a dynamic CKM therefore poses three fundamental questions, namely where to measure, how long existing knowledge remains usable, and how frequently it should be refreshed.

Existing studies largely address these questions from a map-centric perspective. To determine where to measure, active measurement selection for CKM construction targets locations that maximally reduce map reconstruction error or posterior uncertainty~\cite{shrestha2023surveying}. To determine how long knowledge remains valid and how often to refresh it, dynamic CKM schemes typically update the map at a period dictated by the rate of environmental change~\cite{jiang2025interference}. Such criteria assess map quality in isolation, overlooking that a map error affects performance only when it alters a downstream decision. For instance, if a nearby BS provides a substantially stronger link than any visible satellite, even a large estimation error on the satellite links will not alter the association outcome. Conversely, when two candidates offer comparable link quality, a minor error can flip the decision. Partial offloading instead depends only on the serving link, where any error distorts the split ratio. Consequently, the most uncertain map entry is not necessarily the most valuable one to measure. Likewise, the usable lifetime of channel knowledge, which we term the decision coherence time, depends on the decision it supports, so links evolving at the same physical rate may remain usable for minutes in one decision yet expire within seconds in another. Furthermore, in a shared dynamic CKM, one measurement informs decisions across neighboring cells and subsequent slots, including those involving satellites yet to enter view.

We therefore study task-driven maintenance of a dynamic CKM, valuing each measurement by the expected improvement in the network-level decisions that query the map. Because measurements must be prioritized before their readings arrive, the CKM must track its own uncertainty. Accordingly, we model the CKM as a Gaussian belief over environment residuals indexed by geographic cell, BS, and satellite sky direction. It ages between reports and updates via Kalman filtering. Under this belief, decision regret quantifies the expected utility loss from acting on the map rather than the true channel, and both this regret and its expected reduction from a candidate measurement are computable before measuring. We then formulate CKM maintenance as minimizing the long-run regret under a measurement budget. How much a measurement reduces regret and how fast regret grows with age jointly address all three aforementioned questions regarding dynamic CKM maintenance. Because the high-dimensional belief makes exact optimization intractable, we develop task-driven CKM maintenance (TD-CKM), which evaluates each candidate measurement in closed form by its expected regret reduction across current and upcoming decisions and adds measurements to the batch of each slot in order of value until the budget is reached or no value exceeds the price. The main contributions of this paper are summarized as follows.
\begin{itemize}
\item We formulate dynamic CKM maintenance as an active information acquisition problem that minimizes the long-run regret of network-level decisions subject to a measurement budget. By indexing satellite shadowing by sky direction, the proposed CKM allows shadowing knowledge to persist across successive satellite passes.

\item We develop TD-CKM, an efficient maintenance algorithm derived from a Lagrangian relaxation of the measurement budget. Beyond free routine reports, TD-CKM approximates in closed form the expected regret reduction of each candidate measurement across both current decisions and upcoming decisions predicted via the satellite ephemeris, while its adaptive pricing mechanism enforces the long-run measurement budget on every sample path.

\item We analytically characterize measurement value and knowledge lifetime, showing that a measurement is most valuable when two leading candidates have comparable quality, and that the decision coherence time depends on the decision margin and the large-scale coherence time.

\item We validate TD-CKM through city-scale simulations of an LEO-assisted edge network. Under an identical measurement budget, TD-CKM achieves lower decision regret for both user association and computation offloading than map-centric baselines, and matches the regret of the strongest baseline using 30\% to 40\% fewer extra measurements.
\end{itemize}

\section{Related Work}\label{sec:related}

CKMs have enabled environment-aware designs, including hybrid beamforming, trajectory planning for cellular-connected unmanned aerial vehicles (UAVs), predictive routing and resource allocation over dynamic low-altitude networks, and handover~\cite{wu2024ckmbf,zeng2021snarm,li2026routing,echevarria2026handover}. Their construction from sparse measurements has been studied along several lines. The data requirements for CKM construction were quantified in~\cite{xu2024cgm}, physical and virtual scatterer-based models were exploited in~\cite{sun2025scatterer,sun2026virtual} to estimate channel gain maps, and wireless radiation fields were reconstructed with 3D Gaussian splatting in~\cite{wen2026wrfgs}. To decide where to measure, active schemes use the uncertainty of a Gaussian process ensemble to select informative locations or design measurement trajectories for an energy-constrained UAV~\cite{polyzos2024bayesian,chen2026aerialckm}. For time-varying environments, a streaming algorithm reconstructs spatial-spectral-temporal radio maps, while a CKM-based framework senses structural variations of the propagation environment from historical CKM data and real-time user channels~\cite{chen2025dynamicsc,li2026sensing}. The CKM in~\cite{yang2025lsm} represents line-of-sight probabilities as a binary random field updated by Bayesian filtering as measurements arrive. The authors in~\cite{song2026partialckm} integrated the completion of a partial, aging CKM into UAV trajectory optimization, revealing a trade-off between navigation and map completion. Across these studies, measurements are selected and maps refreshed to improve map accuracy. Downstream decisions shape measurement selection only when the measuring agent also decides, as in~\cite{zeng2021snarm,song2026partialckm}.

On the decision side, computation offloading to LEO edge servers, multi-layer offloading among BSs and LEO servers, and joint association and handover between BSs and LEO satellites have been optimized for LEO-assisted edge networks~\cite{tang2021leo,cao2023edge,nguyenkha2025handover}. The design in~\cite{nguyenkha2025handover} predicts link gains over a time window using a three-dimensional city map. The value of fresh information to a decision has been studied mainly for status updates. The age of information (AoI) measures update freshness, whereas the age of incorrect information measures the time spent with an incorrect state estimate~\cite{yates2021aoi,maatouk2020aoii}. For an Ornstein-Uhlenbeck source, error-based sampling attains a lower estimation error than age-based sampling~\cite{ornee2021sampling}. Goal-oriented sampling values updates by their impact on the receiver's task, as in the joint design of sampling and remote decision making~\cite{li2026freshness} and in the semantic-aware tracking of a partially observable source~\cite{zakeri2025semantic}. With a limited budget for several sources, update scheduling becomes a restless bandit problem. Whittle index policies have been derived for general cost functions of age and for remote estimation of Gauss-Markov processes~\cite{tripathi2019whittle,ornee2026whittle}. These index policies value updates to a single receiver, for either a single source or independent sources. By contrast, a CKM entry is correlated with its neighbors and queried by heterogeneous decisions across cells and slots, while satellite motion shifts which entries these decisions query. How to quantify the value of a measurement for decisions that query such a shared, dynamic CKM remains open.

\section{System Model}\label{sec:system}

\subsection{Network Model}\label{subsec:network}

\begin{figure}[t]
\centering
\includegraphics[width=\columnwidth]{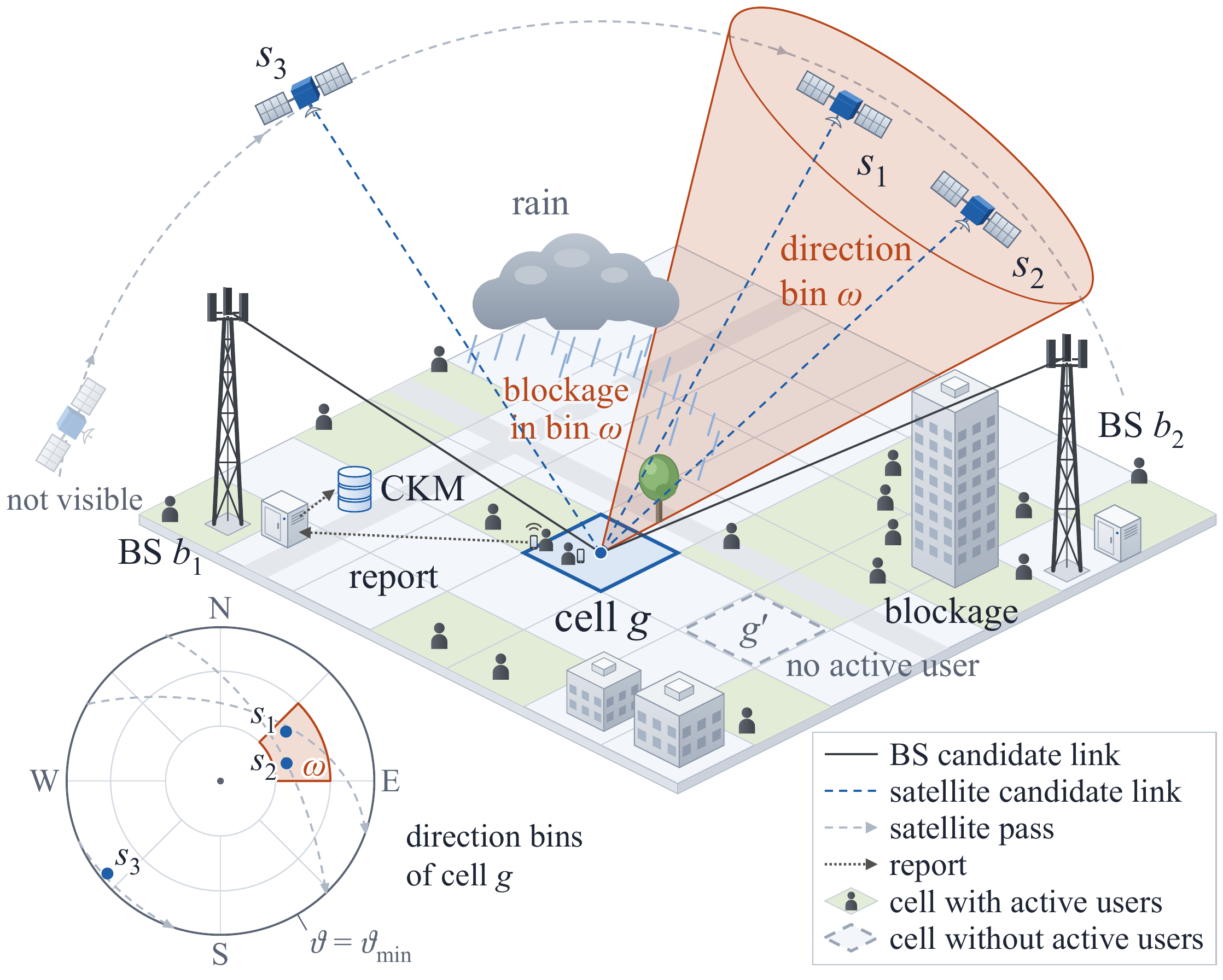}
\caption{An LEO-assisted edge network with a dynamic CKM. Satellites $s_1$ and $s_2$ lie in one direction bin $\omega$ of cell $g$ and share the same shadowing entry. Cell~$g'$ has no active user to report. The inset is a sky plot of the direction bins of cell $g$.}
\label{fig:scene}
\end{figure}

We consider an LEO-assisted edge network, as shown in Fig.~\ref{fig:scene}, where terrestrial BSs with edge servers and LEO satellites with onboard servers provide communication and computing to users in a ground region. A regional CKM server maintains a dynamic CKM for user association and computation offloading decisions. The region is partitioned into $G$ geographic cells, indexed by $g\in\mathcal G=\{1,\dots,G\}$, which form the spatial units of the CKM and are represented by their reference points $\mathbf x_g$. The network operates in slots $t\in\mathcal T=\{1,\dots,H\}$ of duration $\Delta$, with satellite positions evaluated at each slot's start. The terrestrial layer comprises $N_{\rm B}$ BSs, indexed by $b\in\mathcal B=\{1,\dots,N_{\rm B}\}$, and the satellite layer comprises $N_{\rm S}$ LEO satellites, indexed by $s\in\mathcal S=\{N_{\rm B}+1,\dots,N_{\rm B}+N_{\rm S}\}$. Each cell $g$ has a nonempty candidate BS set $\mathcal B_g\subseteq\mathcal B$. Viewed from $\mathbf x_g$, satellite $s$ has an elevation angle $\vartheta_{g,s}(t)$ determined by its ephemeris position $\mathbf p_s(t)$. Satellites above the minimum elevation angle $\vartheta_{\min}$ constitute the visible set $\mathcal S_g(t)=\{s\in\mathcal S:\vartheta_{g,s}(t)\ge\vartheta_{\min}\}$. Cell $g$ has candidate nodes $\mathcal K_g(t)=\mathcal B_g\cup\mathcal S_g(t)$ and candidate links $\mathcal L_g(t)=\{(g,k):k\in\mathcal K_g(t)\}$. Following CKM grid discretization~\cite{zeng2024tutorial}, link $(g,k)$ characterizes the large-scale channel between $\mathbf x_g$ and node $k$, approximating that of all users within cell $g$.

Each connected user is served by one candidate BS or satellite of its cell. Users become active, move, and depart independently of the propagation environment and network decisions, while the network knows the activity statistics and active user count $n_g(t)$ of each cell $g$ at the start of slot $t$.
\subsection{Channel Model}\label{subsec:channel}

For each candidate link $(g,k)\in\mathcal L_g(t)$, let $\zeta_{g,k}(t)$ denote its downlink large-scale SNR in dB. Following the separation of static and dynamic environments in CKMs~\cite{zeng2024tutorial}, $\zeta_{g,k}(t)$ is decomposed into a known mean $\bar\zeta_{g,k}(t)$ and zero-mean environment residuals in dB. Terrestrial BS links operate below 6~GHz with negligible rain attenuation~\cite{tr38811}, whereas Ka-band satellite links experience rain attenuation along slant paths that sweep across the sky as satellites pass. Denoting by $z^{\rm sh}_{g,b}(t)$ and $z^{\rm sh}_{g,s}(t)$ the shadowing residuals toward BS $b$ and along the path to satellite $s$, respectively, and by $z^{\rm r}_g(t)$ the rain residual of cell $g$, the SNRs of BS link $(g,b)$ and satellite link $(g,s)$ are expressed as
\begin{align}
\zeta_{g,b}(t)&=\bar\zeta_{g,b}+z^{\rm sh}_{g,b}(t),\label{eq:bs-link}\\
\zeta_{g,s}(t)&=\bar\zeta_{g,s}(t)+z^{\rm sh}_{g,s}(t)+\alpha_{g,s}(t)\,z^{\rm r}_g(t),\label{eq:sat-link}
\end{align}
where both means contain known link-budget terms, including long-term building and terrain shadowing stored in an offline CKM toward the BS or satellite sky direction, and $\bar\zeta_{g,s}(t)$ further includes gaseous, cloud, and mean rain attenuation~\cite{tr38811}. Treating rain as uniform below the rain height, $z^{\rm r}_g(t)$ represents the rain residual along the zenith path from cell $g$, scaled on a slant path by $\alpha_{g,s}(t)=1/\sin\vartheta_{g,s}(t)$.

Averaged over small-scale fading, the rate of link $(g,k)$ is approximated by
\begin{equation}\label{eq:rate}
R_{g,k}(\zeta_{g,k})=W_k\log_2\big(1+10^{\zeta_{g,k}/10}\big),
\end{equation}
where $W_k$ is the bandwidth allocated by node $k$ to a served user on resources orthogonal to other links. The residuals vary over minutes to hours, slowly relative to slots of seconds~\cite{itup1853}. With the means and slant-path factors known, these residuals are the only link unknowns maintained by the CKM.
\subsection{CKM Model}\label{subsec:ckm}

For BS links, the CKM stores one shadowing entry $z^{\rm sh}_{g,b}(t)$ per cell $g$ and candidate BS $b\in\mathcal B_g$, as in a conventional CKM. By contrast, satellite shadowing is indexed by sky direction, since blockage depends on the propagation path regardless of which satellite is at its far end. As shown in the inset of Fig.~\ref{fig:scene}, the sky above $\vartheta_{\min}$ viewed from $\mathbf x_g$ is partitioned into direction bins $\omega\in\Omega$, each spanning an elevation band and an azimuth sector, and $\omega_{g,s}(t)$ denotes the bin containing the direction from $\mathbf x_g$ to $\mathbf p_s(t)$. All satellites in one bin share the shadowing residual $z^{\rm sh}_{g,s}(t)=z^{\rm sh}_{g,\omega_{g,s}(t)}(t)$, so its knowledge carries over between satellites. They remain separate candidates with their own means and slant-path factors. Only one rain residual $z^{\rm r}_g(t)$ is stored per cell, since $\alpha_{g,s}(t)$ maps it to every slant path.

Stacking $z^{\rm sh}_{g,b}(t)$, $z^{\rm sh}_{g,\omega}(t)$, and $z^{\rm r}_g(t)$ of all cells $g$ in a fixed order over an index set $\mathcal I$ with $N=|\mathcal I|$ yields the environment state $\mathbf z_t\in\mathbb R^N$. With the SNRs and means of all candidate links in slot $t$ stacked into $\boldsymbol\zeta_t$ and $\bar{\boldsymbol\zeta}_t$, the SNRs are given by
\begin{equation}\label{eq:readout}
\boldsymbol\zeta_t=\bar{\boldsymbol\zeta}_t+\mathbf C_t\mathbf z_t,
\end{equation}
where row $\mathbf c_{g,k}^\top(t)$ of $\mathbf C_t$ has a unit coefficient on the shadowing entry of link $(g,k)$ and, for a satellite link $(g,s)$, the slant-path factor $\alpha_{g,s}(t)$ on the rain entry $z^{\rm r}_g$. Satellite motion alters only $\bar{\boldsymbol\zeta}_t$ and $\mathbf C_t$, both known in advance from the ephemeris, so the CKM tracks only $\mathbf z_t$.

Following~\cite{wu2026sameckm}, we represent a CKM as a posterior distribution of channel knowledge and define the CKM in slot $t$ as the Gaussian belief over $\mathbf z_t$ conditioned on all reports $\mathcal D_t$ received up to slot $t$, given by
\begin{equation}\label{eq:ckm}
\mathcal M_t\triangleq p(\mathbf z_t\mid\mathcal D_t)=\Normal(\boldsymbol\mu_t,\boldsymbol\Sigma_t).
\end{equation}
The induced link belief is $\boldsymbol\zeta_t\mid\mathcal D_t\sim\Normal(\bar{\boldsymbol\zeta}_t+\mathbf C_t\boldsymbol\mu_t,\,\mathbf C_t\boldsymbol\Sigma_t\mathbf C_t^\top)$. This belief remains Gaussian because the prior residuals are jointly Gaussian and each report provides a linear observation in Gaussian noise.

The belief is initialized at the stationary prior $\Normal(\mathbf 0,\boldsymbol\Sigma_\infty)$. Shadowing residuals toward the same BS or direction bin, as well as rain residuals, are correlated across cells, with exponential decay in distance~\cite{shrestha2023surveying}. All other pairs of entries are uncorrelated. For $j\in\mathcal B\cup\Omega$ and any two cells $g$ and $g'$, the nonzero entries of $\boldsymbol\Sigma_\infty$ are given by
\begin{equation}\label{eq:prior}
\begin{aligned}
\Cov\big(z^{\rm sh}_{g,j},z^{\rm sh}_{g',j}\big)&=\sigma_{{\rm sh},j}^2\exp\big(-\|\mathbf x_g-\mathbf x_{g'}\|/d_{\rm sh}\big),\\
\Cov\big(z^{\rm r}_g,z^{\rm r}_{g'}\big)&=\sigma_{{\rm r},g}\,\sigma_{{\rm r},g'}\exp\big(-\|\mathbf x_g-\mathbf x_{g'}\|/d_{\rm r}\big),
\end{aligned}
\end{equation}
where $\sigma_{{\rm sh},j}$ and $\sigma_{{\rm r},g}$ denote stationary standard deviations and $d_{\rm sh}$ and $d_{\rm r}\gg d_{\rm sh}$ are decorrelation distances. With entries grouped by BS, direction bin, and rain, $\boldsymbol\Sigma_\infty$ is block diagonal.

The residuals follow a first-order Gauss-Markov model, as in the Kriged Kalman model of~\cite{zeng2024tutorial}, given by
\begin{equation}\label{eq:dynamics}
\mathbf z_{t+1}=\mathbf F\mathbf z_t+\mathbf w_t,\qquad \mathbf w_t\sim\Normal\big(\mathbf 0,\boldsymbol\Sigma_\infty-\mathbf F\boldsymbol\Sigma_\infty\mathbf F\big),
\end{equation}
where $\mathbf F=\diag\big(\exp(-\Delta/T_i)\big)_{i\in\mathcal I}$ and $\mathbf w_t$ is independent across slots and of $\mathbf z_t$. This choice preserves $\Cov(\mathbf z_t)=\boldsymbol\Sigma_\infty$ in every slot. Correlated entries share a time constant, $T_{\rm sh}$ for shadowing and $T_{\rm r}$ for rain, so $\mathbf F$ scales each block of $\boldsymbol\Sigma_\infty$ uniformly, ensuring $\boldsymbol\Sigma_\infty-\mathbf F\boldsymbol\Sigma_\infty\mathbf F=(\mathbf I-\mathbf F^2)\boldsymbol\Sigma_\infty\succeq\mathbf 0$. Each entry $i$ thus follows a stationary Ornstein-Uhlenbeck process sampled once per slot~\cite{ornee2021sampling}, with autocorrelation $\exp(-|t'-t|\Delta/T_i)$ between slots $t$ and $t'$, where $T_i$ is its large-scale coherence time. For rain, the Gaussian residual in dB approximates the log-normal attenuation model of~\cite{itup1853} around its climatological mean. The CKM server knows $\boldsymbol\Sigma_\infty$ and the coherence times, which can be fitted offline to routine reports, and treats $\mathbf z_t$ as constant within a slot since $\Delta\ll T_i$.

Between reports, each entry's marginal belief ages in closed form. If a report leaves entry $i$ with posterior mean $\mu_0$ and variance $\sigma_0^2$ and no report follows, its mean and variance at age $\tau$ are given by
\begin{equation}\label{eq:aging}
\begin{aligned}
\mu(\tau)&=\exp(-\tau/T_i)\,\mu_0,\\
\sigma^2(\tau)&=\sigma_\infty^2-(\sigma_\infty^2-\sigma_0^2)\exp(-2\tau/T_i),
\end{aligned}
\end{equation}
where $\sigma_\infty^2=[\boldsymbol\Sigma_\infty]_{ii}$. The same expressions hold for any linear combination of entries sharing a common coherence time, such as the difference of two BS shadowing entries, with $\mu_0$, $\sigma_0^2$, and $\sigma_\infty^2$ evaluated for that combination.
\subsection{Measurement and Update Model}\label{subsec:measurement}

CKM reports come only from active user terminals, so links in cell $g$ can be measured in slot $t$ only if $n_g(t)\ge1$. These dual-mode terminals, such as vehicle-mounted or fixed devices, can measure both BS and satellite links. For mobility management, connected users report candidate links in the layer of their serving nodes. To limit uplink and backhaul signaling, routine reports follow a periodic schedule and are forwarded upon receipt by the serving nodes to the CKM server. The server ingests all routine reports received in slot $t$, denoted by $\mathcal F_t$, at no extra cost. Any other report requires an extra measurement. Extra measurements fill gaps left by routine reports, including candidates in the other layer, links in cells whose active users are not yet connected, and reported links that blockage or rain may have changed since their last routine report. Each extra measurement consumes terminal energy, incurs signaling overhead, and requires a measurement gap when probing the other layer. With a common unit cost, the candidate reports in slot $t$ are $\mathcal C_t=\{(g,k): n_g(t)\ge1,\ k\in\mathcal K_g(t)\}$. Extra measurements satisfy $\mathcal Q_t\subseteq\mathcal C_t\setminus\mathcal F_t$ and $|\mathcal Q_t|\le B$, where $\mathcal F_t\subseteq\mathcal C_t$ and $B$ is the per-slot network budget.

For a candidate report $q=(g,k)\in\mathcal C_t$, the observation $Y_q$ is the measured large-scale SNR minus the known mean $\bar\zeta_{g,k}(t)$. As a scalar linear observation of $\mathbf z_t$, it satisfies $Y_q=\mathbf c_q^\top\mathbf z_t+\varepsilon_q$, where $\varepsilon_q\sim\Normal(0,\sigma_{\rm n}^2)$. Here $\mathbf c_q\triangleq\mathbf c_{g,k}(t)$, and the noise $\varepsilon_q$, independent of $\mathbf z_t$ and across reports, captures measurement error and the effects of user offsets from $\mathbf x_g$ and satellite direction offsets within a bin.

The CKM server tracks $\mathbf z_t$ with a Kalman filter initialized at the stationary prior. At the start of slot $t$, the predicted mean and covariance are given by
\begin{equation}\label{eq:predict}
\boldsymbol\mu_t^-=\mathbf F\boldsymbol\mu_{t-1},\qquad
\boldsymbol\Sigma_t^-=\mathbf F\boldsymbol\Sigma_{t-1}\mathbf F+\boldsymbol\Sigma_\infty-\mathbf F\boldsymbol\Sigma_\infty\mathbf F.
\end{equation}
Reports are then assimilated sequentially, each updating a belief $\mathcal M=\Normal(\boldsymbol\mu,\boldsymbol\Sigma)$ to $\mathcal M^{+q}=\Normal(\boldsymbol\mu^{+q},\boldsymbol\Sigma^{+q})$ as
\begin{equation}\label{eq:kalman}
\boldsymbol\mu^{+q}=\boldsymbol\mu+\mathbf v_q\,\frac{Y_q-\mathbf c_q^\top\boldsymbol\mu}{\sqrt{S_q}},\qquad
\boldsymbol\Sigma^{+q}=\boldsymbol\Sigma-\mathbf v_q\mathbf v_q^\top,
\end{equation}
where $S_q=\mathbf c_q^\top\boldsymbol\Sigma\mathbf c_q+\sigma_{\rm n}^2$ is the innovation variance of $Y_q$ and $\mathbf v_q=\boldsymbol\Sigma\mathbf c_q/\sqrt{S_q}$ is the update direction. Since $\boldsymbol\Sigma\mathbf c_q=\Cov(\mathbf z_t,Y_q\mid\mathcal M)$, a report updates every entry correlated with $Y_q$. Reports from nearby cells or a second satellite at another elevation in the same bin thus help separate the shadowing and rain components observed jointly by one satellite report.

Before the reading arrives, the standardized innovation $Z_q\triangleq(Y_q-\mathbf c_q^\top\boldsymbol\mu)/\sqrt{S_q}$ follows $\Normal(0,1)$ under $\mathcal M$, so the posterior mean moves along $\mathbf v_q$ according to
\begin{equation}\label{eq:innovation}
\boldsymbol\mu^{+q}=\boldsymbol\mu+\mathbf v_qZ_q,\qquad Z_q\sim\Normal(0,1),
\end{equation}
whereas the covariance loses $\mathbf v_q\mathbf v_q^\top$ regardless of the reading, as in the knowledge gradient~\cite{wang2018finite}. The value of a report depends on $\boldsymbol\mu$ through the decisions, so extra measurements are selected online in each slot.

In slot $t$, the CKM server first predicts the belief and assimilates the routine reports $\mathcal F_t$, yielding $\mathcal M_t^{\mathcal F}=\Normal(\boldsymbol\mu_t^{\mathcal F},\boldsymbol\Sigma_t^{\mathcal F})$. Its only decision is to select the extra measurements $\mathcal Q_t$ within the budget using this belief, before any extra reading is known. The extra readings arrive within the slot and update the CKM to $\mathcal M_t=\Normal(\boldsymbol\mu_t,\boldsymbol\Sigma_t)$, from which the network makes its decisions before the environment evolves to the next slot.

\subsection{Task Model}\label{subsec:task}

We consider two representative network-level tasks that query the CKM in distinct ways, namely discrete user association (Task~I) and continuous computation offloading (Task~II). In Task~I, a user in cell $g$ selects a serving node from $\mathcal K_g(t)$, and the utility of choosing candidate $k$ is expressed as
\begin{equation}\label{eq:assoc-utility}
U^{\rm I}(k;\boldsymbol\zeta)=R_{g,k}(\zeta_{g,k})-c^{\rm HO}\,\Ind[k\ne k^{\rm prev}],
\end{equation}
where $\Ind[\cdot]$ denotes the indicator function and the handover penalty $c^{\rm HO}$ captures the throughput loss averaged over the association period when switching away from the previous serving node $k^{\rm prev}$. Consequently, an association decision depends jointly on all candidate links in $\mathcal L_g(t)$.

In Task~II, a user partitions a computation task with $L$ input bits and $\eta$ CPU cycles per bit between local execution and its serving node $k$~\cite{zhang2026tvt}. Let $x\in[0,1]$ denote the offloading ratio, such that $xL$ bits are offloaded over serving link $(g,k)$ while $(1-x)L$ bits are processed locally. The delays for fully local execution and full offloading are given by
\begin{equation}\label{eq:offload-paths}
\delta^{\rm loc}=\frac{\eta L}{f^{\rm loc}},\qquad \delta^{\rm off}(\zeta_{g,k})=\frac{L}{R_{g,k}}+\frac{\eta L}{f_k},
\end{equation}
where $f^{\rm loc}$ and $f_k$ denote the computing rates (in CPU cycles/s) allocated to the task by the user and node $k$, and $R_{g,k}$ is the uplink rate of link $(g,k)$, evaluated via (\ref{eq:rate}) at $\zeta_{g,k}(t)$ plus a known link-budget offset since uplink and downlink share the same large-scale residuals. Under parallel execution, the task completion delay at split ratio $x$ is given by
\begin{equation}\label{eq:offload-delay}
\delta(x)=\max\big\{(1-x)\,\delta^{\rm loc},\,x\,\delta^{\rm off}(\zeta_{g,k})\big\},
\end{equation}
corresponding to utility $U^{\rm II}(x;\boldsymbol\zeta)=-\delta(x)$. For a known $\zeta_{g,k}$, equalizing the two branches minimizes $\delta(x)$, yielding the optimal split ratio and minimum delay given by
\begin{equation}\label{eq:offload-opt}
x^\star=\frac{\delta^{\rm loc}}{\delta^{\rm loc}+\delta^{\rm off}},\qquad \delta(x^\star)=\frac{\delta^{\rm loc}\,\delta^{\rm off}}{\delta^{\rm loc}+\delta^{\rm off}}.
\end{equation}
The return of the small result and the LEO propagation delay are omitted, as both are short compared with the completion delay of computation-intensive tasks~\cite{tr38811}. Thus, each offloading decision depends solely on the serving link through $1/R_{g,k}$.

Each association or offloading instance is represented as a decision event $e=(t_e,g_e,\theta_e,\xi_e)$ occurring in cell $g_e$ during slot $t_e$, where $\theta_e\in\{\mathrm I,\mathrm{II}\}$ specifies the task type and $\xi_e$ collects known context parameters such as $k^{\rm prev}$ or the task parameters. Events are triggered by user activation, visibility changes in $\mathcal S_g(t)$, periodic association re-evaluation, or task arrivals. The action space $\mathcal A_e$ and utility $U_e(a;\boldsymbol\zeta)$ are given by $\mathcal K_{g_e}(t_e)$ and $U^{\rm I}$ for $\theta_e=\mathrm I$, and by $[0,1]$ and $U^{\rm II}$ for $\theta_e=\mathrm{II}$. The event set $\mathcal E_t$ of slot $t$, together with its contexts, is revealed at the start of slot $t$. Offloading events use the serving node at the start of the slot, while updated associations affect routine reporting from slot $t+1$ onward. Future events of type $\theta$ in cell $g$ are characterized statistically by an event rate $\nu_{g,\theta}$ (in $\mathrm s^{-1}$) and a representative context $\bar\xi_{g,\theta}$.

For each event $e$, the network selects the Bayes-optimal action that maximizes the expected utility under its belief $\mathcal M$ over the residuals in slot $t_e$, given by
\begin{equation}\label{eq:bayes-rule}
a_e^{\mathcal M}\in\argmax_{a\in\mathcal A_e}\ \E\big[U_e(a;\boldsymbol\zeta)\,\big|\,\mathcal M\big],
\end{equation}
where $\boldsymbol\zeta=\boldsymbol\zeta_{t_e}$. In slot $t_e$, the network acts under the updated CKM belief $\mathcal M_{t_e}$, whereas an oracle knowing the true $\boldsymbol\zeta$ would select $a_e^\star(\boldsymbol\zeta)\in\argmax_{a\in\mathcal A_e}U_e(a;\boldsymbol\zeta)$. With the Bayes decision rule fixed, CKM maintenance affects the utility loss relative to $a_e^\star(\boldsymbol\zeta)$ only through $\mathcal M_{t_e}$.

\section{Problem Formulation}\label{sec:formulation}

\subsection{Decision Regret}\label{subsec:regret}

The decision regret of event $e$ under belief $\mathcal M$ quantifies the expected utility loss of the Bayes action $a_e^{\mathcal M}$ relative to the oracle action $a_e^\star(\boldsymbol\zeta)$, and is defined as
\begin{equation}\label{eq:regret}
\begin{aligned}
&\mathcal R_e(\mathcal M)\triangleq\E\big[U_e(a_e^\star(\boldsymbol\zeta);\boldsymbol\zeta)-U_e(a_e^{\mathcal M};\boldsymbol\zeta)\,\big|\,\mathcal M\big]\\
&\quad=\E\Big[\max_{a\in\mathcal A_e}U_e(a;\boldsymbol\zeta)\,\Big|\,\mathcal M\Big]-\max_{a\in\mathcal A_e}\E\Big[U_e(a;\boldsymbol\zeta)\,\Big|\,\mathcal M\Big],
\end{aligned}
\end{equation}
which is the expected value of perfect information (EVPI) for event $e$~\cite{tan2024bayesian}. Because the expectation is taken under $\mathcal M$, the regret can be evaluated directly from the CKM.

This regret equals the probability that $a_e^{\mathcal M}$ is suboptimal multiplied by the conditional mean loss given suboptimality. For association, it vanishes when $a_e^{\mathcal M}$ is optimal almost surely. For offloading, uncertainty about the serving link can make the Bayes split differ from $x^\star$.
\subsection{Main Problem}\label{subsec:p1}
\begin{figure*}[t]
\centering
\includegraphics[width=0.92\textwidth]{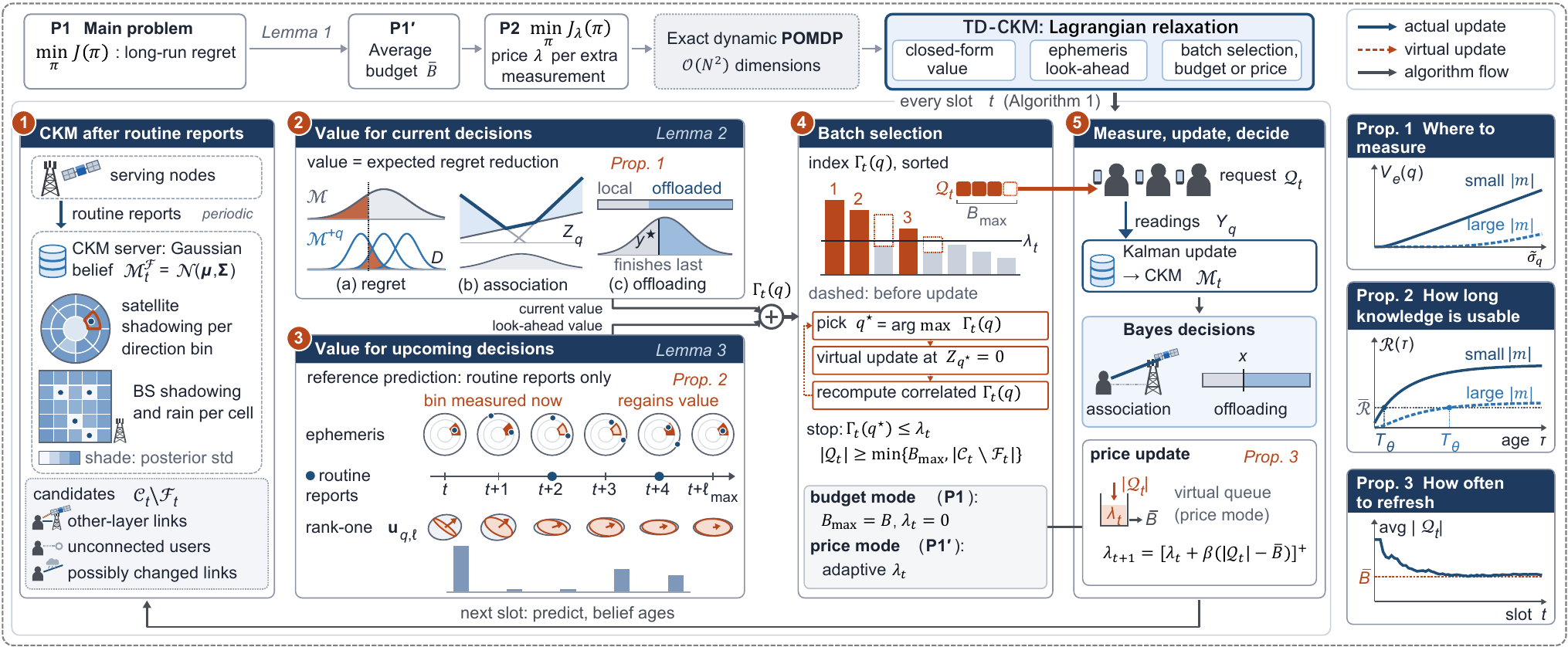}
\caption{Framework of TD-CKM. In each slot, candidate reports are scored and selected via virtual covariance updates before extra readings arrive, after which the readings update the CKM.}
\label{fig:framework}
\end{figure*}

A maintenance policy $\pi=(\pi_t)_{t\ge1}$ selects the extra measurements $\mathcal Q_t$ in each slot $t$. Because a measurement informs decisions in both the current and subsequent slots until its knowledge ages, we evaluate $\pi$ by the long-run average weighted regret, defined as
\begin{equation}\label{eq:p1-cost}
J(\pi)\triangleq\limsup_{H\to\infty}\frac{1}{H}\sum_{t=1}^{H}\E_{\pi}\Big[\sum_{e\in\mathcal E_t}\rho_e\mathcal R_e(\mathcal M_t)\Big],
\end{equation}
where $\rho_e\ge0$ weights event $e$ and $\E_{\pi}[\cdot]$ is taken over the environment dynamics, observation noise, and user activity under $\pi$. The weight reflects event priority and normalizes regret by a reference rate for association or a reference delay for offloading, making the weighted regret dimensionless. The CKM maintenance problem is then formulated as
\begin{subequations}\label{eq:p1}
\begin{alignat}{2}
\Pone\colon\ \ & \min_{\pi}\ \ && J(\pi) \label{eq:p1-obj}\\
& \textbf{s.t.} && \mathcal Q_t\subseteq\mathcal C_t\setminus\mathcal F_t,\quad \forall t, \label{eq:p1-supply}\\
& && |\mathcal Q_t|\le B,\quad \forall t, \label{eq:p1-budget}\\
& && \mathcal Q_t=\pi_t\big(\mathcal D_{t-1},\{Y_q\}_{q\in\mathcal F_t},\mathcal C_t,\mathcal E_t\big),\quad \forall t, \label{eq:p1-causal}
\end{alignat}
\end{subequations}
with optimal value $J_1^\star$. Constraints~(\ref{eq:p1-supply}) and (\ref{eq:p1-budget}) specify the available extra measurements and per-slot budget, respectively. Constraint~(\ref{eq:p1-causal}) enforces causality using past reports, current routine readings, candidate reports, and current events, plus the known ephemeris, forwarding schedule, and activity statistics.

\subsection{Problem Relaxation}\label{subsec:p2}

Because the regret reduction achievable by extra measurements varies across slots, relaxing the hard per-slot budget in \Pone{} to a time-average budget $\bar B$ yields
\begin{equation}\label{eq:p1prime}
\begin{alignedat}{2}
\Ponep\colon\ \ & \min_{\pi}\ \ && J(\pi)\\
& \textbf{s.t.} && \limsup_{H\to\infty}\frac{1}{H}\sum_{t=1}^{H}\E_{\pi}\big[|\mathcal Q_t|\big]\le\bar B,\\
& && \textup{(\ref{eq:p1-supply})},\ \textup{(\ref{eq:p1-causal})},
\end{alignedat}
\end{equation}
which is a relaxation of \Pone{} whenever $\bar B\ge B$. Following the Lagrangian relaxation in restless bandits~\cite{ornee2026whittle}, we dualize the average budget constraint with a price $\lambda\ge0$ per extra measurement and define the Lagrangian cost as
\begin{equation}\label{eq:p2-cost}
J_\lambda(\pi)\triangleq\limsup_{H\to\infty}\frac{1}{H}\sum_{t=1}^{H}\E_{\pi}\Big[\sum_{e\in\mathcal E_t}\rho_e\mathcal R_e(\mathcal M_t)+\lambda|\mathcal Q_t|\Big].
\end{equation}
The resulting problem is formulated as
\begin{equation}\label{eq:p2}
\Ptwo\colon\ \ \min_{\pi}\ \ J_\lambda(\pi)\quad\textbf{s.t.}\ \ \textup{(\ref{eq:p1-supply})},\ \textup{(\ref{eq:p1-causal})}.
\end{equation}
The following lemma establishes their lower-bound ordering.
\begin{lemma}\label{lem:bounds}
For $\bar B=B$, let $J_{1'}^\star$ and $J_2^\star(\lambda)\triangleq\inf_\pi J_\lambda(\pi)$ denote the optimal values of \Ponep{} and \Ptwo{}, respectively. For any $\lambda\ge0$ and any policy $\pi$ feasible for \Pone{}, it holds that
\begin{equation}\label{eq:p2-bounds}
J_2^\star(\lambda)-\lambda\bar B\ \le\ J_{1'}^\star\ \le\ J_1^\star\ \le\ J(\pi).
\end{equation}
\end{lemma}
\begin{proof}
For $\bar B=B$, every policy feasible for \Pone{} is feasible for \Ponep{}, which gives the last two inequalities. For any $\pi'$ feasible for \Ponep{}, $\lambda\ge0$ and the subadditivity of the limit superior give $J_2^\star(\lambda)\le J_\lambda(\pi')\le J(\pi')+\lambda\bar B$, and taking the infimum over $\pi'$ gives the first.
\end{proof}

\section{TD-CKM: Algorithm Design and Analysis}\label{sec:tdckm}

\subsection{Solution Approach}\label{subsec:approach}

Problem \Pone{} is a partially observable Markov decision process (POMDP), whose hidden state $\mathbf z_t$ is observed through routine reports $\mathcal F_t$ and extra measurements $\mathcal Q_t$~\cite{zakeri2025semantic}. Under the linear Gaussian model, the intermediate CKM belief $\mathcal M_t^{\mathcal F}$, together with the slot index, active cells, serving nodes, and current events $\mathcal E_t$, forms a sufficient statistic for dynamic programming. Exact optimization is intractable because measurements inform later decisions as knowledge ages, coupling actions across slots through a belief with $\mathcal O(N^2)$ continuous dimensions. Moreover, selecting up to $B$ reports from thousands of candidates yields a combinatorial action space.

Under \Ptwo{}, each extra measurement is charged a price $\lambda$. Its value combines current weighted regret reduction with the reduction in the optimal future cost. Since computing that cost requires the same dynamic program, TD-CKM evaluates it under a base policy with no further extra measurements. Future measurement charges then vanish, leaving only weighted regret costs. TD-CKM approximates their reduction over the current slot and the next $\ell_{\max}$ slots along a reference path where routine reports continue.

In each slot, TD-CKM selects the report with the greatest current marginal value, virtually updates the covariance, and recomputes remaining values until capacity is reached or no value exceeds the price. Only the selected batch $\mathcal Q_t$ is measured, and valuation repeats in the next slot from the updated belief. The values govern where to measure and account for how long knowledge remains useful, while the budget or price governs how often it is refreshed. Fig.~\ref{fig:framework} summarizes TD-CKM.

\subsection{Measurement Value for Current Decisions}\label{subsec:value}

Let $\mathcal M=\Normal(\boldsymbol\mu,\boldsymbol\Sigma)$ be the working belief, initialized at $\mathcal M_t^{\mathcal F}$ in slot $t$. The value of a candidate report $q\in\mathcal C_t\setminus\mathcal F_t$ for event $e$ is the expected value of sample information (EVSI) of its reading $Y_q$~\cite{tan2024bayesian}, defined as
\begin{equation}\label{eq:evsi}
\begin{aligned}
V_e(q)\triangleq{}&\E_{Y_q}\Big[\max_{a\in\mathcal A_e}\E\big[U_e(a;\boldsymbol\zeta)\mid\mathcal M^{+q}\big]\Big]\\
&-\max_{a\in\mathcal A_e}\E\big[U_e(a;\boldsymbol\zeta)\mid\mathcal M\big],
\end{aligned}
\end{equation}
where $\mathcal M^{+q}$ is the posterior belief obtained by (\ref{eq:kalman}) with $Y_q\sim\Normal(\mathbf c_q^\top\boldsymbol\mu,S_q)$.

\begin{lemma}\label{lem:evsi}
If $\E\big[\max_{a\in\mathcal A_e}|U_e(a;\boldsymbol\zeta)|\mid\mathcal M\big]<\infty$, the EVSI $V_e(q)$ equals the expected reduction in decision regret (\ref{eq:regret}), given by
\begin{equation}\label{eq:value-1d}
V_e(q)=\mathcal R_e(\mathcal M)-\int_{-\infty}^{\infty}\mathcal R_e\big(\mathcal M^{+q}(z)\big)\,\phi(z)\,{\rm d}z,
\end{equation}
where $\phi$ is the standard normal density and $\mathcal M^{+q}(z)=\Normal(\boldsymbol\mu+\mathbf v_qz,\,\boldsymbol\Sigma-\mathbf v_q\mathbf v_q^{\top})$ is the posterior belief at $Z_q=z$. Moreover, $0\le V_e(q)\le\mathcal R_e(\mathcal M)$.
\end{lemma}
\begin{proof}
By (\ref{eq:regret}) and the tower property, the first term of (\ref{eq:evsi}) equals $\E[\max_{a}U_e(a;\boldsymbol\zeta)\mid\mathcal M]-\E_{Y_q}[\mathcal R_e(\mathcal M^{+q})]$, which gives (\ref{eq:value-1d}). Since each expected utility is unchanged on average, Jensen's inequality gives $V_e(q)\ge0$, and $\mathcal R_e(\mathcal M^{+q})\ge0$ gives $V_e(q)\le\mathcal R_e(\mathcal M)$~\cite{tan2024bayesian}.
\end{proof}

Although the scalar innovation $Z_q$ makes (\ref{eq:value-1d}) a one-dimensional integral, each quadrature point still requires recomputing the expected candidate utilities or the Bayes split. Evaluating every report-event pair and future slot is costly, so TD-CKM uses closed-form approximations. Association benefits when a reading can change the winning candidate, whereas offloading benefits by refining its continuous split.

\subsubsection{Association}\label{subsubsec:value-assoc}
For an association event $e$ in cell $g$, let $\bar u_{e,k}=\E[U_e(k;\boldsymbol\zeta)\mid\mathcal M]$ denote the current expected utility of candidate $k\in\mathcal K_g(t)$. By (\ref{eq:innovation}), report $q$ shifts the SNR mean $\hat\zeta_{g,k}=\bar\zeta_{g,k}(t)+\mathbf c_{g,k}^\top\boldsymbol\mu$ by $\mathbf c_{g,k}^\top\mathbf v_qZ_q$. Linearizing $R_{g,k}$ at $\hat\zeta_{g,k}$ with derivative $R'_{g,k}$ gives the affine approximation
\begin{equation}\label{eq:assoc-affine}
\E\big[U_e(k;\boldsymbol\zeta)\mid\mathcal M^{+q}(Z_q)\big]\approx\bar u_{e,k}+\tilde\sigma_{k,q}Z_q,
\end{equation}
where $\tilde\sigma_{k,q}=R'_{g,k}\,\mathbf c_{g,k}^\top\mathbf v_q$ is the utility sensitivity to $Z_q$. Substituting (\ref{eq:assoc-affine}) into (\ref{eq:evsi}) approximates $V_e(q)$ by the expected upper envelope minus $\max_k\bar u_{e,k}$. Let the $\tilde K$ envelope lines have increasing slopes $\tilde\sigma_{(1)}<\cdots<\tilde\sigma_{(\tilde K)}$, with adjacent intersections $y_i=(\bar u_{(i)}-\bar u_{(i+1)})/(\tilde\sigma_{(i+1)}-\tilde\sigma_{(i)})$. Let $(i^\star)$ index the largest intercept, breaking ties by smaller slope, so it attains the envelope at $Z_q=0$. Relative to this line, the envelope is expressed as
\begin{equation}\label{eq:envelope-decomp}
\begin{aligned}
&\max_{k}\big(\bar u_{e,k}+\tilde\sigma_{k,q}Z_q\big)=\bar u_{(i^\star)}+\tilde\sigma_{(i^\star)}Z_q\\
&\quad+\sum_{i=1}^{\tilde K-1}\big(\tilde\sigma_{(i+1)}-\tilde\sigma_{(i)}\big)\big[s_i(Z_q-y_i)\big]^+,
\end{aligned}
\end{equation}
where $[x]^+=\max\{x,0\}$, $s_i=1$ for $y_i\ge0$, and $s_i=-1$ otherwise. Define $\psi(y)\triangleq\phi(y)-y[1-\Phi(y)]=\E[(Z_q-y)^+]$, where $\Phi$ is the standard normal distribution function. Since $\E[Z_q]=0$ and $\E\big[[s_i(Z_q-y_i)]^+\big]=\psi(|y_i|)$, taking expectations in (\ref{eq:envelope-decomp}) and subtracting $\bar u_{(i^\star)}$ yields the closed-form knowledge gradient~\cite{wang2018finite}, given by
\begin{equation}\label{eq:envelope}
V_e(q)\approx\sum_{i=1}^{\tilde K-1}\big(\tilde\sigma_{(i+1)}-\tilde\sigma_{(i)}\big)\,\psi\big(|y_i|\big).
\end{equation}

\subsubsection{Offloading}\label{subsubsec:value-offload}
For an offloading event over serving link $(g,k)$, the network computes the Bayes split by one-dimensional convex optimization. For measurement valuation, we linearize $\delta^{\rm off}$ around $\hat\zeta_{g,k}$ as $\delta^{\rm off}\approx\mu^{\rm off}+\sigma^{\rm off}Z$, where $Z\sim\Normal(0,1)$ and $\mu^{\rm off}=\E[\delta^{\rm off}(\zeta_{g,k})\mid\mathcal M]$. With $R_{g,k}$ and $R'_{g,k}$ evaluated at $\hat\zeta_{g,k}$, the linearized delay standard deviations before and after $q$ are given by
\begin{equation}\label{eq:offload-std}
\sigma^{\rm off}=\frac{LR'_{g,k}}{R_{g,k}^2}\big(\mathbf c_{g,k}^\top\boldsymbol\Sigma\mathbf c_{g,k}\big)^{1/2},\quad
\sigma^{\rm off}_{+q}=\sqrt{(\sigma^{\rm off})^2-(\tilde\sigma^{\rm off}_q)^2},
\end{equation}
where $\tilde\sigma^{\rm off}_q=(LR'_{g,k}/R_{g,k}^2)\,|\mathbf c_{g,k}^\top\mathbf v_q|$ is the magnitude of the delay sensitivity to $Z_q$. Thus, $q$ informs offloading through the projection $\mathbf c_{g,k}^\top\mathbf v_q$ onto the serving link. Let $\bar x=\delta^{\rm loc}/(\delta^{\rm loc}+\mu^{\rm off})$ balance the two branches at the mean delay. For $x>0$, parameterize the split by the standardized delay $y$ satisfying $(1-x)\delta^{\rm loc}=x(\mu^{\rm off}+\sigma^{\rm off}y)$. To first order in the linearized serving-rate coefficient of variation $\sigma^{\rm off}R_{g,k}/L$, the Bayes split is characterized by the classical newsvendor problem, given by
\begin{equation}\label{eq:offload-quantile}
\begin{aligned}
y^\star&=\argmin_{y}\,\E\big[(1-\bar x)(Z-y)^++\bar x\,(y-Z)^+\big]\\
&=\Phi^{-1}(1-\bar x),
\end{aligned}
\end{equation}
where the two terms penalize the offloaded and local branches, respectively, for finishing last. Since the expected penalty equals $\psi(y)+\bar xy$, its minimum is $\phi(y^\star)$, giving the leading-order regret $\bar x\,\phi(y^\star)\,\sigma^{\rm off}$. Report $q$ reduces the standard deviation to $\sigma^{\rm off}_{+q}$. Averaged over its reading, the coefficient $\bar x\,\phi(y^\star)$ changes only at second order, so Lemma~\ref{lem:evsi} gives the value. Since full local execution guarantees $\delta^{\rm loc}$ and the oracle delay is nonnegative, we cap both expressions at $\delta^{\rm loc}$, giving
\begin{equation}\label{eq:offload-value}
\begin{aligned}
\mathcal R_e(\mathcal M)&\approx\min\big\{\bar x\,\phi(y^\star)\,\sigma^{\rm off},\,\delta^{\rm loc}\big\},\\
V_e(q)&\approx\min\big\{\bar x\,\phi(y^\star)\big(\sigma^{\rm off}-\sigma^{\rm off}_{+q}\big),\,\delta^{\rm loc}\big\}.
\end{aligned}
\end{equation}
\subsection{Measurement Value for Upcoming Decisions}\label{subsec:lookahead}

An extra measurement in slot~$t$ also informs later decisions, with a value governed by belief aging, future satellite visibility, and subsequent routine reports. TD-CKM evaluates these gains along a common reference path over $\ell_{\max}$ future slots.

\subsubsection{Reference Prediction}\label{subsubsec:la-reference}
For each lag $\ell\in\{1,\dots,\ell_{\max}\}$, the ephemeris specifies candidates $\mathcal K_g(t+\ell)$, readout rows $\mathbf c_{g,k}(t+\ell)$, and known means $\bar\zeta_{g,k}(t+\ell)$. Other inputs follow a reference path without extra measurements. Active users remain in their cells over the window. Each connected user retains its serving node while it remains a candidate and otherwise re-associates by the Bayes action under the predicted belief. The predicted routine reports $\hat{\mathcal F}_{t+\ell}$ cover candidate links in the layers of serving nodes scheduled to forward in slot $t+\ell$. Future events of type $\theta$ in cell $g$ are represented by event rate $\nu_{g,\theta}$, representative context $\bar\xi_{g,\theta}$ with the reference serving node, and weight $\rho_{g,\theta}$.

The reference belief $\hat{\mathcal M}_{t+\ell\mid t}=\Normal(\hat{\boldsymbol\mu}_{t+\ell\mid t},\hat{\boldsymbol\Sigma}_{t+\ell\mid t})$ propagates $\mathcal M_t^{\mathcal F}=\Normal(\boldsymbol\mu,\boldsymbol\Sigma)$ along this path. Setting future routine readings to their predicted values gives $\hat{\boldsymbol\mu}_{t+\ell\mid t}=\mathbf F^{\ell}\boldsymbol\mu$. Starting from $\hat{\boldsymbol\Sigma}_{t\mid t}=\boldsymbol\Sigma$, the covariance follows the time update (\ref{eq:predict}) and the measurement update (\ref{eq:kalman}) for every $q'\in\hat{\mathcal F}_{t+\ell}$. The initial reference prediction is computed once per slot and stores the innovation variance $S_{q'}$ and update direction $\mathbf v_{q'}$ of each $q'$.

\subsubsection{Rank-One Propagation}\label{subsubsec:la-propagation}
Recomputing the reference covariance for each candidate costs $\mathcal O(N^2)$ per time or routine-report update. Given the stored reference quantities, TD-CKM propagates each candidate's effect through one vector at an additional cost of $\mathcal O(N)$ per update, as follows.

\begin{lemma}\label{lem:lookahead}
Let $\hat{\boldsymbol\Sigma}^{+q}_{t+\ell\mid t}$ denote the reference covariance with $q$ measured in slot~$t$ on the same reference path, initialized at $\boldsymbol\Sigma-\mathbf v_q\mathbf v_q^{\top}$. For $\ell=0,\dots,\ell_{\max}$, the two reference covariances satisfy
\begin{equation}\label{eq:la-rankone}
\hat{\boldsymbol\Sigma}_{t+\ell\mid t}-\hat{\boldsymbol\Sigma}^{+q}_{t+\ell\mid t}=\mathbf u_{q,\ell}\mathbf u_{q,\ell}^{\top},
\end{equation}
where $\mathbf u_{q,0}=\mathbf v_q$, and $\mathbf u_{q,\ell}$ is obtained from $\mathbf u=\mathbf F\mathbf u_{q,\ell-1}$ by sequentially applying
\begin{equation}\label{eq:la-recursion}
\mathbf u\leftarrow\sqrt{\frac{S_{q'}}{S_{q'}-h_{q'}^2}}\left(\mathbf u-\frac{h_{q'}}{\sqrt{S_{q'}}}\,\mathbf v_{q'}\right),\qquad h_{q'}=\mathbf c_{q'}^{\top}\mathbf u,
\end{equation}
for each $q'\in\hat{\mathcal F}_{t+\ell}$ in the reference assimilation order. Here, $S_{q'}-h_{q'}^2\ge\sigma_{\rm n}^2$ is the innovation variance of $q'$ given $q$. Moreover, before observing $Y_q$, its update to the mean of $\mathbf z_{t+\ell}$ conditional on routine readings through slot $t+\ell$ is $\mathbf u_{q,\ell}Z$, where $Z\sim\Normal(0,1)$ is independent of those routine readings.
\end{lemma}
\begin{proof}
At $\ell=0$, the difference is $\mathbf v_q\mathbf v_q^\top$ by (\ref{eq:kalman}). Since the noise term of (\ref{eq:predict}) does not depend on $q$, each time update maps a difference $\mathbf u\mathbf u^\top$ to $(\mathbf F\mathbf u)(\mathbf F\mathbf u)^\top$. For a routine report $q'$, subtracting the updates (\ref{eq:kalman}) with innovation variances $S_{q'}$ and $S_{q'}-h_{q'}^2$ leaves a rank-one difference with factor (\ref{eq:la-recursion}). The mean update follows from the independence of the standardized residual of $Y_q$ from the routine readings.
\end{proof}

By Lemma~\ref{lem:lookahead}, measuring $q$ acts on the belief at slot $t+\ell$ as a single observation with update direction $\mathbf u_{q,\ell}$. Since the ephemeris determines the future readout rows $\mathbf c_{g,k}(t+\ell)$ onto which it is projected, a measured bin can regain value when a later satellite enters it.

\subsubsection{Look-Ahead Index}\label{subsubsec:la-index}
As in (\ref{eq:value-1d}), define the value of $q$ for a representative event $(t+\ell,g,\theta,\bar\xi_{g,\theta})$ as its expected regret reduction on the reference prediction,
\begin{equation}\label{eq:future-value}
\begin{aligned}
V_{g,\theta,\ell}(q)\triangleq{}&\mathcal R_{g,\theta,\ell}\big(\hat{\mathcal M}_{t+\ell\mid t}\big)\\
&-\E_Z\Big[\mathcal R_{g,\theta,\ell}\big(\hat{\mathcal M}^{+q}_{t+\ell\mid t}(Z)\big)\Big],
\end{aligned}
\end{equation}
where $\mathcal R_{g,\theta,\ell}$ is the regret (\ref{eq:regret}) of this event and $\hat{\mathcal M}^{+q}_{t+\ell\mid t}(Z)=\Normal(\hat{\boldsymbol\mu}_{t+\ell\mid t}+\mathbf u_{q,\ell}Z,\,\hat{\boldsymbol\Sigma}^{+q}_{t+\ell\mid t})$. Substituting $\hat{\mathcal M}_{t+\ell\mid t}$, direction $\mathbf u_{q,\ell}$, and readout rows $\mathbf c_{g,k}(t+\ell)$ into (\ref{eq:envelope}) and (\ref{eq:offload-value}) approximates this value in closed form. Weighting future events by their expected count $\nu_{g,\theta}\Delta$ per slot yields the index
\begin{equation}\label{eq:index}
\begin{aligned}
\Gamma_t(q)\triangleq{}&\sum_{e\in\mathcal E_t}\rho_e V_e(q)\\
&+\sum_{\ell=1}^{\ell_{\max}}\sum_{g\in\mathcal G}\sum_{\theta\in\{\mathrm I,\mathrm{II}\}}\nu_{g,\theta}\Delta\,\rho_{g,\theta}\,V_{g,\theta,\ell}(q).
\end{aligned}
\end{equation}
Each closed-form value is bounded by a multiple of $\max_k|\mathbf c_{g,k}^\top(t+\ell)\mathbf u_{q,\ell}|$, so TD-CKM restricts the spatial sum in (\ref{eq:index}) to cells $\mathcal G_q$ with non-negligible projections.

\subsection{Batch Selection and Budget Control}\label{subsec:selection}

\subsubsection{Sequential Selection}\label{subsubsec:sel-greedy}
Virtual covariance updates can change decision uncertainties and the value of further reports, so TD-CKM constructs $\mathcal Q_t$ sequentially. Starting from $\mathcal M=\mathcal M_t^{\mathcal F}$, each step selects the candidate $q^\star$ with the largest current index. Since $Y_{q^\star}$ is unknown, $q^\star$ is virtually assimilated at $Z_{q^\star}=0$, leaving $\boldsymbol\mu$ and $\hat{\boldsymbol\mu}_{t+\ell\mid t}$ unchanged. By (\ref{eq:kalman}) and Lemma~\ref{lem:lookahead}, the covariances are updated as
\begin{equation}\label{eq:greedy-update}
\boldsymbol\Sigma\leftarrow\boldsymbol\Sigma-\mathbf v_{q^\star}\mathbf v_{q^\star}^{\top},\quad
\hat{\boldsymbol\Sigma}_{t+\ell\mid t}\leftarrow\hat{\boldsymbol\Sigma}_{t+\ell\mid t}-\mathbf u_{q^\star,\ell}\mathbf u_{q^\star,\ell}^{\top}
\end{equation}
for $\ell=1,\dots,\ell_{\max}$. Since (\ref{eq:greedy-update}) is rank one, the stored reference quantities $(S_{q'},\mathbf v_{q'})$ are updated in $\mathcal O(N)$ per routine report. The resulting indices approximate marginal values beyond the selected reports. TD-CKM recomputes indices affected through the current belief or reference prediction.

\subsubsection{Budget and Price Modes}\label{subsubsec:sel-modes}
Selection stops when the batch reaches capacity $B_{\max}$ or no remaining index exceeds $\lambda_t$. For \Pone{}, the budget mode sets $B_{\max}=B$ and $\lambda_t=0$. For \Ponep{}, the price mode uses a configurable per-slot capacity $B_{\max}\ge\bar B$ and updates the price at the end of each slot as
\begin{equation}\label{eq:price-update}
\lambda_{t+1}=\big[\lambda_t+\beta\big(|\mathcal Q_t|-\bar B\big)\big]^+,
\end{equation}
where $\beta>0$ is the step size and $\lambda_1\ge0$ is the initial price. Thus $\lambda_t/\beta$ is a virtual queue that accumulates the extra measurements beyond $\bar B$~\cite{bi2021lyapunov}, and its quadratic Lyapunov function $\lambda_t^2/(2\beta)$ satisfies the one-slot drift bound
\begin{equation}\label{eq:price-drift}
\frac{\lambda_{t+1}^2-\lambda_t^2}{2\beta}\le\beta C+\lambda_t\big(|\mathcal Q_t|-\bar B\big),
\end{equation}
where $C=\frac12\max\{\bar B^2,(B_{\max}-\bar B)^2\}$. With indices fixed, minimizing this bound minus $\sum_{q\in\mathcal Q_t}\Gamma_t(q)$ is the drift-plus-penalty rule with trade-off parameter $1/\beta$, which selects the largest indices above $\lambda_t$ up to capacity. TD-CKM applies this threshold one report at a time, updating the indices after each selection. Algorithm~\ref{alg:tdckm} summarizes TD-CKM.

\begin{algorithm}[t]
\caption{TD-CKM}\label{alg:tdckm}
\KwInput{Window $\ell_{\max}$, budget $B$ (budget mode) or capacity $B_{\max}$, average budget $\bar B$, step size $\beta$, and initial price $\lambda_1\ge0$ (price mode)}
\KwOutput{Extra measurements $\mathcal Q_t$ and CKM $\mathcal M_t$}
Initialize $\mathcal M_0\leftarrow\Normal(\mathbf 0,\boldsymbol\Sigma_\infty)$. In budget mode, set $B_{\max}\leftarrow B$ and $\lambda_t\leftarrow0$ for all $t$\;
\For{$t=1$ \KwTo $H$}{
Perform time update (\ref{eq:predict}), observe $\mathcal C_t$, $\mathcal F_t$, and $\mathcal E_t$, assimilate $\mathcal F_t$ via (\ref{eq:kalman}) to obtain $\mathcal M_t^{\mathcal F}$, and initialize $\mathcal M\leftarrow\mathcal M_t^{\mathcal F}$\;
Compute the reference prediction and store $(S_{q'},\mathbf v_{q'})$ over the window\;
Evaluate $\Gamma_t(q)$ via (\ref{eq:index}) and (\ref{eq:la-recursion}) for all $q\in\mathcal C_t\setminus\mathcal F_t$, and initialize $\mathcal Q_t\leftarrow\emptyset$\;
\While{$|\mathcal Q_t|<\min\{B_{\max},|\mathcal C_t\setminus\mathcal F_t|\}$}{
$q^\star\leftarrow\argmax_{q\in\mathcal C_t\setminus(\mathcal F_t\cup\mathcal Q_t)}\Gamma_t(q)$\;
\lIf{$\Gamma_t(q^\star)\le\lambda_t$}{\textbf{break}}
Add $q^\star$ to $\mathcal Q_t$, apply (\ref{eq:greedy-update}), update the stored quantities, and recompute affected indices\;
}
Request $\mathcal Q_t$, assimilate its readings into the saved $\mathcal M_t^{\mathcal F}$ to obtain $\mathcal M_t$, and execute decisions $e\in\mathcal E_t$ via (\ref{eq:bayes-rule})\;
In price mode, update $\lambda_{t+1}$ via (\ref{eq:price-update})\;
}
\end{algorithm}

Per slot, the reference prediction costs $\mathcal O\big(N^2(1+\ell_{\max}+|\hat{\mathcal F}|)\big)$ for the $|\hat{\mathcal F}|$ routine reports in the window, while the virtual updates of the selection cost $\mathcal O\big(B_{\max}N(N(1+\ell_{\max})+|\hat{\mathcal F}|)\big)$. Each index requires $\mathcal O\big(N(1+\ell_{\max}+|\hat{\mathcal F}|)\big)$ for direction propagation, plus one closed-form evaluation per relevant current event or representative future event.

\subsection{Theoretical Analysis}\label{subsec:analysis}

In a two-action model, Propositions~\ref{prop:two} and~\ref{prop:coherence} address where to measure and how long knowledge remains usable. Proposition~\ref{prop:price} establishes TD-CKM's sample-path budget feasibility.

\begin{proposition}[Measurement Value]\label{prop:two}
Let a two-action event $e$ have utility margin $D\triangleq U_e(1;\boldsymbol\zeta)-U_e(2;\boldsymbol\zeta)=m+\boldsymbol\gamma^\top(\mathbf z_t-\boldsymbol\mu)$ under $\mathcal M=\Normal(\boldsymbol\mu,\boldsymbol\Sigma)$, so $D\mid\mathcal M\sim\Normal(m,\sigma^2)$ with $\sigma^2=\boldsymbol\gamma^\top\boldsymbol\Sigma\boldsymbol\gamma$. Its regret and the EVSI of report $q$ are given by
\begin{align}
\mathcal R_e(\mathcal M)&=\sigma\,\psi\big(|m|/\sigma\big),\label{eq:two-regret}\\
V_e(q)&=\tilde\sigma_q\,\psi\big(|m|/\tilde\sigma_q\big),\quad \tilde\sigma_q=|\boldsymbol\gamma^\top\mathbf v_q|\le\sigma,\label{eq:two-value}
\end{align}
where $y\,\psi(|m|/y)$ is defined as $0$ at $y=0$, increases strictly in $y\ge0$, and decreases strictly in $|m|$ for $y>0$.
\end{proposition}
\begin{proof}
Since $U_e(1;\boldsymbol\zeta)=U_e(2;\boldsymbol\zeta)+D$, the Bayes action and the oracle exceed the expected utility of action~2 by $\max\{m,0\}$ and $\E[\max\{D,0\}]=\max\{m,0\}+\sigma\psi(|m|/\sigma)$, respectively, which gives (\ref{eq:two-regret}). Report $q$ shifts the margin mean to $m+\tilde\sigma_qZ_q$, so (\ref{eq:two-value}) follows from (\ref{eq:evsi}) in the same way and is the two-alternative knowledge gradient~\cite{wang2018finite}. The monotonicity follows from $\partial[y\psi(c/y)]/\partial y=\phi(c/y)>0$ and $\partial[y\psi(c/y)]/\partial c=\Phi(c/y)-1<0$, where $c=|m|$.
\end{proof}

At fixed $m$, (\ref{eq:two-value}) grows with the removed margin variance $\tilde\sigma_q^2$, and at fixed $\tilde\sigma_q$ it peaks at $m=0$. Rankings by map variance or mutual information ignore $m$ and $\boldsymbol\gamma$ and can therefore misorder measurements.

The regret (\ref{eq:two-regret}) depends on the belief only through the margin mean $m$ and standard deviation $\sigma$. To isolate knowledge aging, hold $m$ fixed and assume a common coherence time $T$ for its residuals, with no further reports. After a report leaves standard deviation $\sigma_0$, the variance $\sigma^2(\tau)$ follows (\ref{eq:aging}) with $T_i=T$ and $\sigma_\infty^2=\boldsymbol\gamma^\top\boldsymbol\Sigma_\infty\boldsymbol\gamma$. The regret at age $\tau$ is given by
\begin{equation}\label{eq:coh-regret}
\mathcal R(\tau)=\sigma(\tau)\,\psi\big(|m|/\sigma(\tau)\big),
\end{equation}
which approaches $\mathcal R_\infty=\sigma_\infty\psi(|m|/\sigma_\infty)$ as knowledge ages.

\begin{definition}\label{def:coherence}
For a regret tolerance $\bar{\mathcal R}>0$, define the decision coherence time as $T_\theta\triangleq\inf\{\tau\ge0:\mathcal R(\tau)\ge\bar{\mathcal R}\}$, with $\inf\emptyset=\infty$.
\end{definition}

\begin{proposition}[Decision Coherence Time]\label{prop:coherence}
Under (\ref{eq:coh-regret}) with $T>0$ and $0<\sigma_0<\sigma_\infty$, the decision coherence time is $T_\theta=0$ if $\bar{\mathcal R}\le\mathcal R(0)$ and $T_\theta=\infty$ if $\bar{\mathcal R}\ge\mathcal R_\infty$. Otherwise, it is given by
\begin{equation}\label{eq:tcoh}
T_\theta=\frac{T}{2}\ln\frac{\sigma_\infty^2-\sigma_0^2}{\sigma_\infty^2-\sigma_{\bar{\mathcal R}}^2},
\end{equation}
where $\sigma_{\bar{\mathcal R}}\in(\sigma_0,\sigma_\infty)$ uniquely solves $\sigma\psi(|m|/\sigma)=\bar{\mathcal R}$. Moreover, $T_\theta$ is nondecreasing in both $|m|$ and $\bar{\mathcal R}$.
\end{proposition}
By (\ref{eq:aging}) and Proposition~\ref{prop:two}, $\mathcal R(\tau)$ increases with $\tau$ from $\mathcal R(0)$ toward $\mathcal R_\infty$, which gives the boundary cases. Otherwise, (\ref{eq:tcoh}) follows from setting $\sigma(\tau)=\sigma_{\bar{\mathcal R}}$ in (\ref{eq:aging}) and solving for $\tau$. Since $\sigma_{\bar{\mathcal R}}$ increases with both $|m|$ and $\bar{\mathcal R}$, so does $T_\theta$.

\begin{proposition}[Budget Feasibility]\label{prop:price}
Let $\beta>0$, $\lambda_1\ge0$, and $0\le\bar B\le B_{\max}$. Suppose all indices evaluated in the price mode satisfy $\Gamma_t(q)\le\Gamma_{\max}<\infty$. Then, on every sample path, $0\le\lambda_t\le\lambda^{\rm ub}\triangleq\max\{\lambda_1,\Gamma_{\max}+\beta(B_{\max}-\bar B)\}$ for all $t\ge1$, and the average extra measurements over any horizon $H$ satisfy
\begin{equation}\label{eq:price-average}
\frac{1}{H}\sum_{t=1}^{H}|\mathcal Q_t|\le\bar B+\frac{\lambda^{\rm ub}-\lambda_1}{\beta H},
\end{equation}
so the price mode is feasible for \Ponep{}.
\end{proposition}
\begin{proof}
If $\lambda_t\ge\Gamma_{\max}$, no index exceeds the price, so $\mathcal Q_t=\emptyset$ and $\lambda_{t+1}\le\lambda_t$. Otherwise, $\lambda_{t+1}<\Gamma_{\max}+\beta(B_{\max}-\bar B)$, so induction gives $0\le\lambda_t\le\lambda^{\rm ub}$. Summing $\lambda_{t+1}\ge\lambda_t+\beta(|\mathcal Q_t|-\bar B)$ over $t\le H$ and using $\lambda_{H+1}\le\lambda^{\rm ub}$ yields (\ref{eq:price-average}).
\end{proof}

\section{Simulation Results}\label{sec:sim}

\subsection{Simulation Setup}\label{subsec:sim-setup}

\begin{figure}[t]
\centering
\includegraphics[width=\columnwidth]{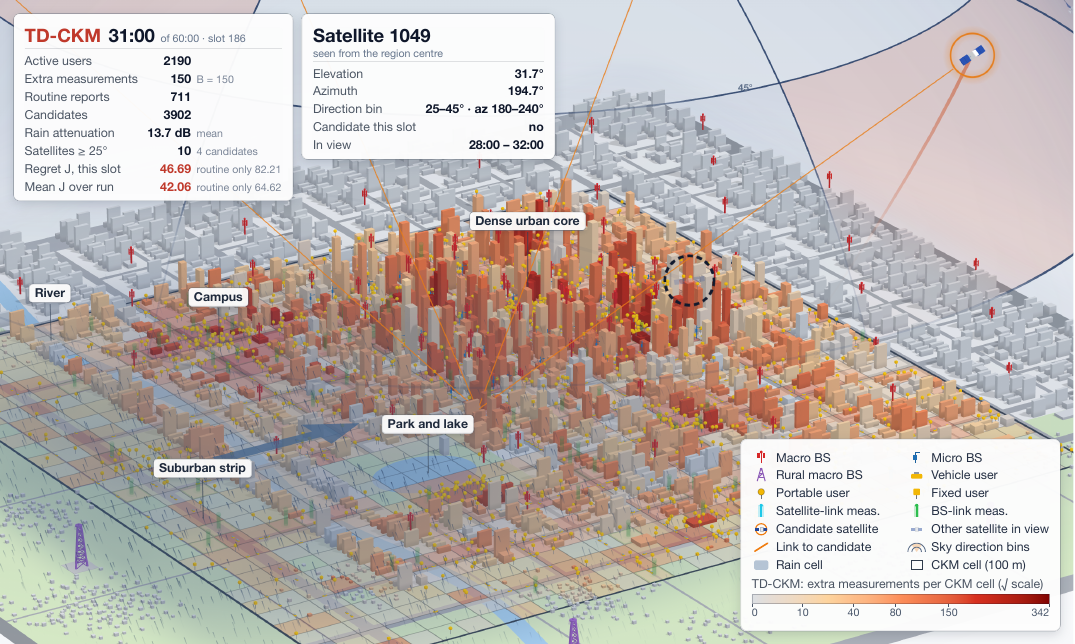}
\caption{3D view of the simulated district at the 31st minute of a run with $B=150$. Cells and buildings are shaded by the number of extra measurements that TD-CKM takes in each cell over the run.}
\label{fig:sim-scene}
\end{figure}

We build a 3D simulation scenario of a 3~km $\times$ 3~km urban district, as shown in Fig.~\ref{fig:sim-scene}. The district comprises $G=900$ square cells of side length 100~m, covered by 89 macro BSs, 64 micro BSs, and a Walker-delta LEO shell. Terrestrial links follow the 3GPP urban macro, urban micro, and rural macro models, while satellite links follow the 3GPP non-terrestrial network model in~\cite{nguyenkha2024twotier}. Unless otherwise stated, each cell considers three BSs and up to four satellites above $\vartheta_{\min}=25^\circ$ as candidates and indexes satellite shadowing by 13 sky direction bins. About 2,600 dual-mode terminals are active on average, mounted on vehicles, carried by pedestrians near hotspots and on a campus, or installed at fixed locations.

To test robustness, the simulated channels include effects beyond the CKM model, namely random blockages of 8~dB to 16~dB over several cells for a few minutes, intra-bin shadowing variations, and user-specific deviations from the representative cell link. Each run also includes a moving rain cell and a campus arrival surge. CKM statistics are fitted offline to residuals from 5 training runs. Serving nodes forward routine reports every 30~s on a staggered schedule, yielding about 1,050 routine reports among 4,700 candidate reports per slot of $\Delta=10$~s. Active users re-evaluate association synchronously every 60~s and start offloading tasks every 30~s on average. Table~\ref{tab:sim-params} lists the main parameters.

TD-CKM and its baselines share the same belief model, candidates, routine reports, and budget, differing only in how they select $\mathcal Q_t$. We compare TD-CKM with two decision-aware and three map-centric baselines.
\begin{itemize}
\item \textbf{Regret-priority}: Ranks decisions by their regret (\ref{eq:regret}) and sequentially measures the link most likely to alter each decision~\cite{wang2025mef}.
\item \textbf{Demand-weighted}: Selects reports that most reduce the variances of links queried by current and upcoming decisions, weighted by link demand and rate sensitivity~\cite{wu2026probe}.
\item \textbf{Max-variance}: Selects reports with the largest predicted variance~\cite{shrestha2023surveying}.
\item \textbf{MSE greedy}: Selects reports that most reduce the CKM mean squared error (MSE), measured by its total posterior variance~\cite{chamon2021approximate}.
\item \textbf{AoI}: Prioritizes entries with the largest AoI~\cite{yates2021aoi}.
\end{itemize}
TD-CKM, Demand-weighted, Max-variance, and MSE greedy virtually update the covariance after each selection. Two~reference policies are also evaluated, namely Routine only, which takes no extra measurements, and All candidates, which measures every candidate link not covered by routine reports.

We evaluate policies by $\hat J$, the mean weighted slot regret over evaluation slots. Realized regret is computed from true SNRs unobserved by the policies. Under matched CKM statistics, expected realized regret equals the expectation of (\ref{eq:regret}) by the tower property, so $\hat J$ estimates $J(\pi)$ in (\ref{eq:p1-cost}). We also report rate loss per association decision and excess delay per offloading task. Each setting uses 10 seeds, or 5 for mismatch experiments, with independent environments, users, and report noise. We report means and 95\% confidence intervals across seeds.

\begin{table}[t]
\caption{Main Simulation Parameters}
\label{tab:sim-params}
\centering
\footnotesize
\begin{tabular}{@{}ll@{}}
\toprule
Parameter & Value\\
\midrule
BS and satellite carriers (GHz) & 3.5, 20\\
Bandwidth per user $W_k$ (MHz) & 10\\
LEO planes $\times$ satellites per plane & 72 $\times$ 22\\
LEO altitude (km), inclination & 550, 53$^\circ$\\
Cells $G$, cell side (m) & 900, 100\\
Slot duration $\Delta$ (s), run length (min) & 10, 60\\
Direction bins per cell, $\vartheta_{\min}$ & 13, 25$^\circ$\\
Forwarding, re-evaluation periods (s) & 30, 60\\
Handover cost $c^{\rm HO}$ (Mbps) & 1\\
Task input $L$ (Mbit), intensity $\eta$ (cycles/bit) & 0.3--6, 150--1500\\
Local computing rate $f^{\rm loc}$ (GHz) & 1--2\\
Computing rate $f_k$ of a BS, a satellite (GHz) & 5, 2.5\\
Residual std. $\sigma_{{\rm sh},j}$, $\sigma_{{\rm r},g}$ (dB) & 3.3--5.1, 4.5\\
Decorrelation distances $d_{\rm sh}$ (m), $d_{\rm r}$ (km) & 52, 27\\
Coherence times $T_{\rm sh}$, $T_{\rm r}$ (s) & 178, 266\\
Report noise $\sigma_{\rm n}$ (dB) & 1.5\\
Reference rate (Mbps) and delay (s) in $\rho_e$ & 27, 0.2\\
Budget $B$ (per slot) & 25, 50, 100, 150, 200, 300\\
Average budget $\bar B$, capacity $B_{\max}$ (per slot) & 150, 600\\
Step size $\beta$, initial price $\lambda_1$ & $3.52\times10^{-5}$, 0.033\\
Look-ahead window $\ell_{\max}$ (slots) & 1\\
\bottomrule
\end{tabular}
\end{table}

\subsection{Overall Performance}\label{subsec:sim-overall}

\begin{figure*}[t]
\centering
\includegraphics[width=0.92\textwidth]{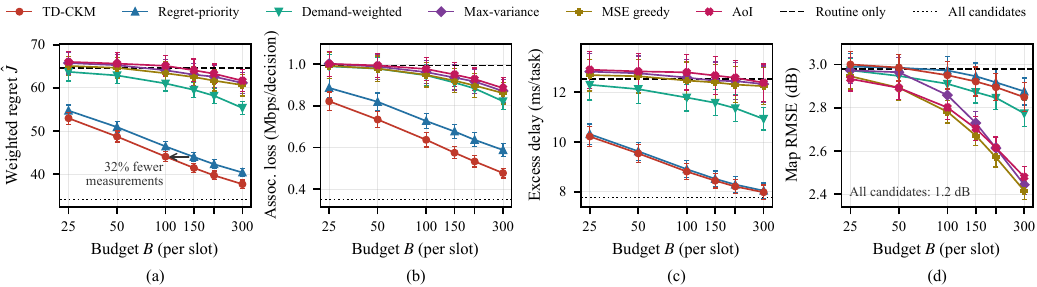}
\caption{Performance versus the budget $B$, with 95\% confidence intervals. The arrow in (a) marks the budget at which TD-CKM attains the regret of Regret-priority with $B=150$, interpolated over $\log B$.}
\label{fig:overall}
\end{figure*}

As shown in Fig.~\ref{fig:overall}(a), TD-CKM achieves the lowest regret across all tested budgets and seeds. Its lead over Regret-priority, the strongest baseline, grows from 3.2\% at $B=25$ to 6.7\% at $B=300$. For the same regret, TD-CKM requires 30\% to 40\% fewer extra measurements. At $B=150$, it measures 4\% of the candidates remaining after routine reports and closes 76\% of the regret gap between Routine only and All candidates. Its association rate loss is 15\% lower than Regret-priority's, while their excess offloading delays differ by about 1\%, as shown in Fig.~\ref{fig:overall}(b) and (c). Both policies refresh the serving links used for offloading. Association compares multiple BS and satellite candidates, making its regret more sensitive to measurement selection. With an adaptive price and $\bar B=150$, TD-CKM averages 150 extra measurements per slot and lowers $\hat J$ by a further 3.3\% relative to budget mode, allocating more measurements to synchronous association re-evaluation slots.

In Fig.~\ref{fig:overall}(d), MSE greedy, AoI, and Max-variance reduce CKM root-mean-square error (RMSE) by 0.25~dB to 0.31~dB at $B=150$ but lower $\hat J$ by less than 4\% relative to Routine only. TD-CKM reduces RMSE by only 0.06~dB yet lowers $\hat J$ by 36\%. RMSE weights all candidate links equally, whereas regret reflects how channel errors affect decisions.

\subsection{Where to Measure}\label{subsec:sim-where}

\begin{figure*}[t]
\centering
\includegraphics[width=0.92\textwidth]{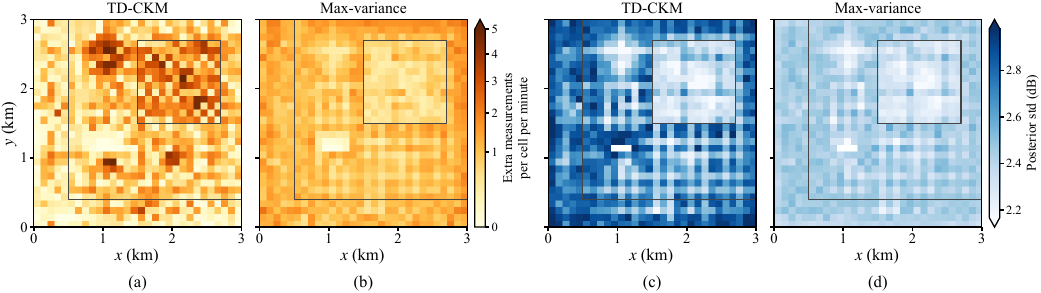}
\caption{Extra measurements and root-mean-square posterior standard deviation of the CKM entries in each cell at $B=150$, averaged over the evaluation slots. The square marks the dense urban core, and the line the edge of the suburban strip.}
\label{fig:maps}
\end{figure*}

\begin{table}[t]
\caption{Performance and Measured Links at $B=150$}
\label{tab:where}
\centering
\footnotesize
\setlength{\tabcolsep}{4pt}
\begin{tabular}{@{}lccccc@{}}
\toprule
 & & Assoc. & Excess & Serving & Satellite\\
Policy & $\hat J$ & loss & delay & links & links\\
 & & (Mbps) & (ms) & (\%) & (\%)\\
\midrule
Routine only & 64.6 & 0.99 & 12.5 & -- & --\\
All candidates & 34.1 & 0.35 & 7.8 & 15 & 61\\
\midrule
TD-CKM & \textbf{41.4} & 0.58 & 8.4 & 64 & 47\\
\quad Association only & 48.9 & \textbf{0.52} & 11.0 & 26 & 58\\
\quad Offloading only & 47.4 & 0.88 & \textbf{8.2} & 98 & 36\\
\midrule
Regret-priority & 44.0 & 0.68 & 8.5 & 71 & 45\\
Demand-weighted & 59.5 & 0.91 & 11.6 & 33 & 88\\
Max-variance & 63.1 & 0.93 & 12.5 & 4 & 66\\
MSE greedy & 62.5 & 0.92 & 12.4 & 5 & 66\\
AoI & 64.1 & 0.95 & 12.7 & 4 & 92\\
\bottomrule
\end{tabular}
\end{table}

TD-CKM concentrates extra measurements where decisions are frequent, as shown in Fig.~\ref{fig:maps}(a) and (b). The dense urban core accounts for 39\% of decisions and receives 30\% of TD-CKM's extra measurements, compared with 9\% under Max-variance. Max-variance spreads its budget across the district because dense routine reporting already keeps uncertainty low in the core. It yields lower and more uniform posterior uncertainty in Fig.~\ref{fig:maps}(c) and (d), with a root-mean-square posterior standard deviation of 2.40~dB versus 2.59~dB for TD-CKM. This lower uncertainty, however, does not improve decisions in high-demand cells. Across the 10\% of cells with the most decisions, TD-CKM reduces the regret of Routine only by 35\%, whereas Max-variance yields no reduction.

TD-CKM and Regret-priority allocate most measurements to serving links, as shown in Table~\ref{tab:where}. These links account for 15\% of the remaining candidates but receive 64\% of TD-CKM's measurements and 71\% of Regret-priority's, versus at most 5\% under map-centric baselines. Routine reports keep their variance and age low, so map-centric criteria favor satellite links. TD-CKM instead targets the decision regret remaining between the 30~s routine reports.

Table~\ref{tab:where} also compares two single-task variants of TD-CKM. The offloading-only variant allocates 98\% of measurements to serving links because each offloading decision depends solely on its serving link. The association-only variant favors alternative BSs and satellites competing with the serving node. Each variant performs best on its own task and worse on the other. TD-CKM considers both tasks and lowers $\hat J$ by 13\% to 15\% relative to either variant.

\subsection{How Long Knowledge Remains Usable}\label{subsec:sim-lifetime}

\begin{figure}[t]
\centering
\begin{subfigure}{0.49\columnwidth}
\centering
\includegraphics[width=\linewidth]{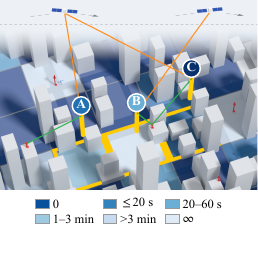}
\caption{$T_\theta$ of each cell}
\label{fig:lifetime-city}
\end{subfigure}\hfill
\begin{subfigure}{0.49\columnwidth}
\centering
\includegraphics[width=\linewidth]{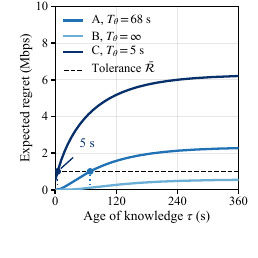}
\caption{Regret against knowledge age}
\label{fig:lifetime-regret}
\end{subfigure}
\caption{Knowledge lifetime in three neighboring cells at the 31st minute of the run in Fig.~\ref{fig:sim-scene}, with $T=178$~s and $\bar{\mathcal R}=c^{\rm HO}$.}
\label{fig:lifetime}
\end{figure}

\begin{figure*}[t]
\centering
\includegraphics[width=0.92\textwidth]{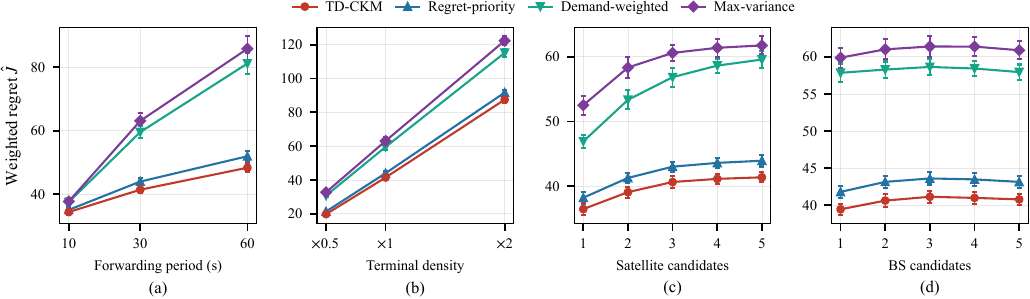}
\caption{Weighted regret at $B=150$ versus forwarding period, terminal density, and candidate satellite and BS counts.}
\label{fig:sweep}
\end{figure*}

\begin{figure}[t]
\centering
\includegraphics[width=0.92\columnwidth]{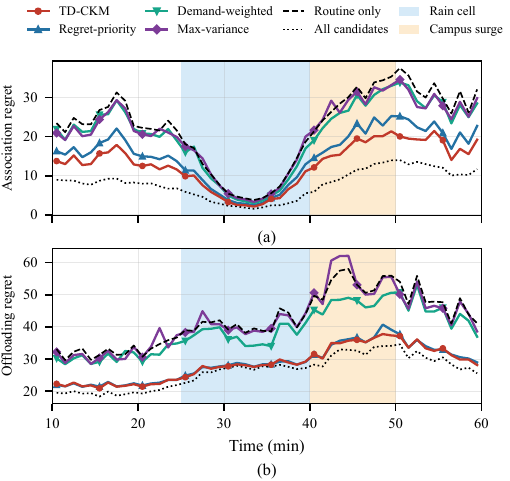}
\caption{Weighted regret per slot of (a) association and (b) offloading at $B=150$, averaged over each minute.}
\label{fig:time}
\end{figure}

\begin{table}[t]
\caption{Weighted Regret Under Mismatched CKM Statistics at $B=150$}
\label{tab:mismatch}
\centering
\footnotesize
\setlength{\tabcolsep}{4pt}
\begin{tabular}{@{}lcccc@{}}
\toprule
 & \multicolumn{2}{c}{$\hat J$} & Lead & Mean\\
CKM statistics & TD-CKM & Regret-priority & (\%) & of $Z_q^2$\\
\midrule
Matched & 32.1 & 35.2 & 8.7 & 1.00\\
Coherence times $\times0.3$ & 35.0 & 38.5 & 9.3 & 0.74\\
Decorrelation distances $\times4$ & 35.3 & 38.5 & 8.3 & 1.29\\
Report noise $\times0.5$ & 36.4 & 39.3 & 7.3 & 1.53\\
Residual std.\ $\times0.5$ & 41.2 & 43.3 & 4.7 & 2.13\\
\bottomrule
\end{tabular}
\end{table}

Knowledge lifetime depends on the querying decision, as illustrated in Fig.~\ref{fig:lifetime} by neighboring cells A, B, and C with the same coherence time and nearly equal margin uncertainty. The same BS has a margin of $m=39.1$~Mbps over the satellite in cell B versus $m=5.6$~Mbps in cell C. At the specified regret tolerance, $T_\theta=\infty$ in B, 5~s in C, and 68~s in A. Larger margins tolerate more uncertainty before regret reaches the tolerance. Across association re-evaluation slots, 53\% of active cells have $T_\theta<20$~s and 17\% have $T_\theta=\infty$. TD-CKM takes about 16 times as many extra measurements per cell for $T_\theta<20$~s as for $T_\theta=\infty$. Map-centric baselines allocate at least as many measurements per cell to the latter. Storing satellite shadowing by sky direction lets knowledge persist beyond individual satellite passes. Under per-satellite storage, where each direction bin is reinitialized from the prior whenever a different satellite enters it, $\hat J$ of TD-CKM at $B=150$ rises by 1.2\% on average and on 9 of 10 seeds.

\subsection{Robustness}\label{subsec:sim-robust}

TD-CKM outperforms Regret-priority on every seed across Fig.~\ref{fig:sweep}(a) and (b), with mean regret 2.0\% to 7.3\% lower. Its lead grows from 2.0\% to 6.9\% as the forwarding period increases from 10~s to 60~s, and narrows from 7.3\% to 4.6\% as terminal density increases fourfold. Less frequent routine reports increase the value of extra measurements, while higher density spreads the same budget over more decisions. Across candidate satellite and BS counts in Fig.~\ref{fig:sweep}(c) and (d), TD-CKM achieves the lowest regret at every count and on every seed, with a 4.6\% to 5.9\% lead over Regret-priority.

Before, during, and after the rain and campus arrival events in Fig.~\ref{fig:time}, TD-CKM has 13\% to 16\% lower association regret than Regret-priority and slightly lower offloading regret. Rain attenuates satellite links by about 8~dB on average, widening margins in favor of BSs and reducing association regret across policies. During the campus surge, TD-CKM increases its campus measurement allocation from 8\% to 17\%, whereas Max-variance reduces its share from 6\% to 4\%.

TD-CKM retains its advantage under CKM statistical mismatch in Table~\ref{tab:mismatch}. Channels follow the CKM model, while one group of policy-assumed statistics is rescaled at a time. Mismatch raises TD-CKM's mean regret by 9\% to 28\% relative to its matched case, but it remains 4.7\% to 9.3\% below Regret-priority under the same mismatch. The mean squared standardized innovation $Z_q^2$ from routine reports shifts from 1.00 with matched statistics to 0.74--2.13 under mismatch, indicating that routine reports alone can flag CKM miscalibration.

Under matched CKM statistics, regret predicted before measurement closely matches mean realized regret. Realized regret exceeds the prediction by 2\% for association and 4\% for offloading, rising to 26\% and 21\%, respectively, when channels include effects beyond the CKM model. All policy comparisons use realized regret.

\section{Conclusion}\label{sec:conclusion}
This paper has studied task-driven maintenance of dynamic CKMs in LEO-assisted edge networks. TD-CKM values extra measurements in closed form by their expected regret reduction and selects them in order of value under a budget or an adaptive price. The analysis shows that measurement value peaks when two candidates have comparable quality, and that knowledge lifetime depends on the decision margin and the large-scale coherence time. City-scale simulations show that TD-CKM matches the strongest baseline's regret with 30\% to 40\% fewer extra measurements, whereas lowering map error alone barely reduces regret. Future work will extend TD-CKM to re-estimate the CKM statistics online from routine-report innovations.

\bibliographystyle{IEEEtran}
\bibliography{IEEEabrv,Ref}

\end{document}